\documentclass{article}
\usepackage{iclr2027_conference}
\iclrfinalcopy 
\usepackage{iftex}
\ifXeTeX
  \usepackage{fontspec}
\else
  \usepackage{times}
\fi

\usepackage{amsmath,amsfonts,bm}

\def\eqref#1{equation~\ref{#1}}

\def\1{\bm{1}}

\DeclareMathAlphabet{\mathsfit}{\encodingdefault}{\sfdefault}{m}{sl}
\SetMathAlphabet{\mathsfit}{bold}{\encodingdefault}{\sfdefault}{bx}{n}

\usepackage{hyperref}
\usepackage{url}
\usepackage{amsmath,amssymb}
\usepackage{graphicx,booktabs,tabularx,array,xcolor,float}
\usepackage[font=small,labelfont=bf]{caption}
\usepackage{enumitem}
\usepackage[most]{tcolorbox}
\newcolumntype{L}{>{\raggedright\arraybackslash}X}

\newcommand{\Rw}{\ensuremath{R_{\mathrm{warn}}}}
\newcommand{\phitwo}{\ensuremath{\varphi_{\geq 2}}}
\newcommand{\rhopop}{\ensuremath{\eta_{\mathrm{pop}}}}
\newcommand{\eown}{\ensuremath{e_{\mathrm{own}}}}
\newcommand{\Rnot}{\ensuremath{R_0}}
\newcommand{\mainfig}[2]{\IfFileExists{#1}{\includegraphics[width=\linewidth]{#1}}{\fbox{\parbox[c][#2][c]{0.96\linewidth}{\centering\small Figure in preparation (spec in STORYBOARD.md):\\ \texttt{\detokenize{#1}}}}}}

\title{Scam Town: Can Scam Education\\ Spread Like a Vaccine?}

\author{\parbox[t]{\dimexpr\textwidth-2\tabcolsep\relax}{\raggedright
Simiao Ren\thanks{All authors contributed equally. Corresponding author: \texttt{benren@scam.ai}.}, Tommy Duong, Xingyu Shen, Jiaqi Gan, Tsang Ng, Ankit Raj, Kidus Zewde, Huaiyu Guo, Siyu Zhang\\[3pt]
\normalfont Scam.ai}}

\setcitestyle{numbers,square,comma}

\begin{document}
\maketitle
{\renewcommand{\thefootnote}{}\footnote{Authorship of this paper does not imply employment by or affiliation with Scam.AI (Reality Inc.); some authors are independent researchers who collaborated on this work.}\addtocounter{footnote}{-1}}

\lhead{Preprint}
\clubpenalty=300 \widowpenalty=300

\begin{abstract}
Most older adults who hear a warning about scams hear it from someone they know: a relative, a neighbour, or a friend who was targeted. Yet research on scam prevention almost always studies one person at a time. We ask whether scam awareness can spread through a community and protect it the way a vaccine does, since a warning protects whoever hears it and can be passed on. To find out, we built \emph{Scam Town}, a simulated community in which a large language model plays each resident. Scam centres phone the residents, and after a call a resident can warn the people they know. Word of mouth helped, but it did not take off on its own: each warning led to fewer than one more, so a community cannot count on warnings ``going viral''. Across 80 simulated towns, everyday talk cut the number of people scammed by roughly a fifth. The largest gains came when an official warning, from a bank, a senior centre, or an agency, reached most of a town and residents then talked about it: in simulation (a combination we did not test live) the two together cut scams by about half. Protection was greatest when warnings reached the people scammers actually call, who here were older residents. Because our residents are language-model stand-ins rather than real people, we treat these findings as predictions to test with people, not as proof. They point to a simple recipe for scam education: give people a clear story they can retell, design campaigns to be passed along, and reach first those most likely to be called.
\end{abstract}

\begin{tcolorbox}[colback=gray!6,colframe=black!55,boxrule=0.5pt,arc=1pt,left=4pt,right=4pt,top=3pt,bottom=3pt,title={\small\textbf{Key takeaways}},fonttitle=\small,coltitle=black,colbacktitle=gray!18]
\small
\begin{enumerate}[leftmargin=1.3em,itemsep=1pt,topsep=0pt]
\item \textbf{Word of mouth cut victimisation in live LLM towns (tested).} In 20 pairs of live LLM towns at survey-level settings, it cut victimisation by 21\% (95\% CI 11--29\%; $-3$ to 40\% with batch noise), as forecast (19\%); when people told more, the cut beat its forecast, so the model is not forecast-validated (\S\ref{sec:live}, Fig.~\ref{fig:live}).
\item \textbf{A vaccine that does not sustain itself, but stretches every dose.} At survey-level settings each warning led to fewer than one more ($\Rw \leq 0.68$ across the literature range), yet in simulation relay let a seeding campaign reach the same effect with about a quarter fewer seeds (\S\ref{sec:phase}, Fig.~\ref{fig:phase}).
\item \textbf{What a warning says matters more than its length (exploratory).} In an exploratory LLM panel, a second account of the scam cut compliance from 12\% to 4\%; padding one warning to the same length did not (14\%), and one neutrally worded warning cut it to 5\% (\S\ref{sec:dose}, Fig.~\ref{fig:dose}).
\item \textbf{A broadcast does far more when people talk about it (simulation only, not tested live).} A trusted broadcast reaching 90\% of a town averted 23\% of victimisations alone and 52\% with word of mouth, under the LLM-measured curve with all believing and never forgetting; talk added 14--29 points under every curve (17--22 if half believe it), but the super-additive part reverses under the in-town curve used for the live forecast (\S\ref{sec:broadcast}, Fig.~\ref{fig:broadcast}).
\item \textbf{Warn the people scammers call, not the best-connected (simulation, targeting assumed).} Seeding 20\% of the town among residents aged 65+ added 11.4 points, among the best-connected 4.6; a broadcast aimed at them first reached a 25\% cut with 15\% of the town warned (random: 25\%; \S\ref{sec:broadcast}).
\end{enumerate}
\end{tcolorbox}

\begin{figure}[t]
\centering
\mainfig{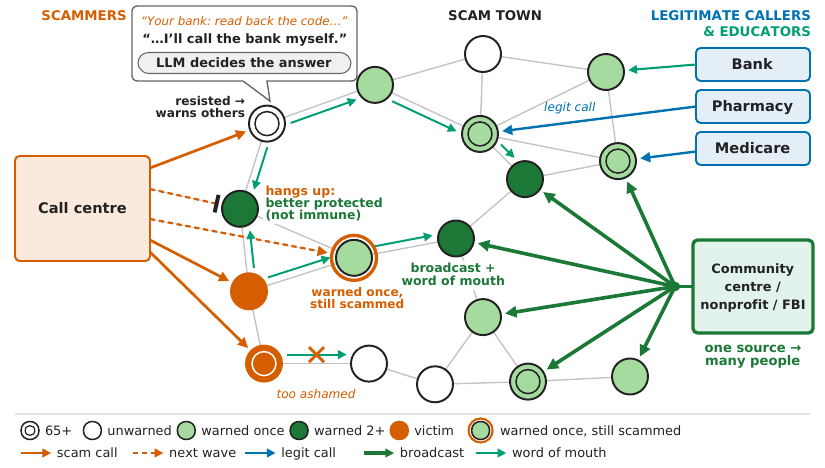}{3.2in}
\caption{\textbf{Scam Town: scam education modelled as a transmissible vaccine (schematic).} A scam call centre (left) and legitimate callers (right: bank, pharmacy, Medicare) phone residents (double circle: aged 65+). A trusted educator (a community centre, a nonprofit, or an agency such as the FBI) broadcasts a tale that warns many residents at once; tales also pass one to one, though some victims stay silent. The LLM decides each answer and the wording of each warning; the model sets ties, calls, who is told and broadcasts. Schematic.}
\label{fig:overview}
\end{figure}

\section{Introduction}

Phone scams are a large and increasingly automated threat: callers now work from scripted playbooks and often use synthetic voices \citep{traister2026anatomy,shen2026machines}. Scam education is usually studied one person at a time \citep{scheibe2014forewarning,burke2022can,deliema2024fraud}. Yet older adults get most scam advice from friends and family \citep{button2024disseminating}, people who meet a scam tell cautionary stories \citep{das2014effect,rader2012stories}, and banks, churches and senior centres add warnings of their own. A tale protects its hearer and can be passed on, so scam education can be modelled as a \emph{transmissible vaccine}, which spreads from host to host and multiplies coverage even when weakly transmissible \citep{nuismer2016eradicating,basinski2019little}. How far do tales spread, how much do they protect, and where should campaigns plant them? No survey reports how a person holding $D$ cautionary tales answers a scam call, so we measure that on LLM residents and model what LLMs do badly (who tells whom, who believes it) from surveys or stated assumptions (Table~\ref{tab:split}). Figure~\ref{fig:overview} shows the resulting \emph{Scam Town}.

The paper has four parts: a model of scam education as an epidemic of cautionary tales set off by scam calls, campaigns and broadcasts (\S\ref{sec:model}); a dose-response of warnings measured on LLM residents, with control arms (\S\ref{sec:dose}); a fast simulator that maps how much talk adds to a broadcast and where education could be aimed, built from a mix of measured, survey-level and assumed parameters (\S\ref{sec:phase}, \S\ref{sec:broadcast}); and a pre-declared live test of that simulator in 80 live LLM towns (\S\ref{sec:live}). Every finding concerns LLM residents. We read them as hypotheses for field trials, not as evidence about people.

\section{Related work}
\label{sec:related}
We treat education itself as the contagion. The closest formal ancestor is the disease--awareness model of \citet{funk2009spread}, in which awareness spreads, fades and lowers susceptibility, but there the threat is also contagious; here, as in ring vaccination, most tales are set off by the threat itself. Inoculation theory treats forewarning as a decaying vaccine against persuasion that can spread through talk \citep{mcguire1964inducing,compton2009spreading,pilditch2022psychological}. Fraud epidemic models make the fraud the contagion \citep{zhu2019investor,guo2026enhancing}. Our seeding and targeting results sit alongside network models of rumour spreading \citep{moreno2004dynamics,nekovee2007theory}, complex contagion, where adoption needs reinforcement from several contacts \citep{dodds2004universal,centola2007complex,centola2010spread}, and degree- and acquaintance-targeted immunisation \citep{pastorsatorras2002immunization,cohen2003efficient}; in all of those the pathogen or rumour itself spreads, whereas here only the protection does. LLM-agent simulations \citep{park2023generative} often stop at believability \citep{williams2023epidemic,larooij2025do}, and LLM stand-ins for people caricature and under-disperse \citep{bisbee2024synthetic,cheng2023compost}. Benchmarks of small language models as scam-call screeners and phone secretaries offer one way to check such call decisions against held-out data \citep{gan2026callscreenbench,ren2026openjev}. We could not calibrate the LLM's call answers to human compliance rates; two attempts failed held-out validation (Appendix~\ref{app:lessons}). We therefore let the LLM decide only what we can check for plausibility, and model the rest. One structural feature sets our setting apart. In every transmissible-vaccine model we surveyed, the vaccine spreads through a population in which the pathogen also spreads; here the threat does not spread at all, and only the protection does. To our knowledge, no prior work measures how an agent's scam response changes with the number of cautionary tales it holds. The nearest human evidence compares six mailed warnings with one \citep{deliema2024fraud}.

\section{Scam Town: the model}
\label{sec:model}

\begin{figure}[t]
\centering
\mainfig{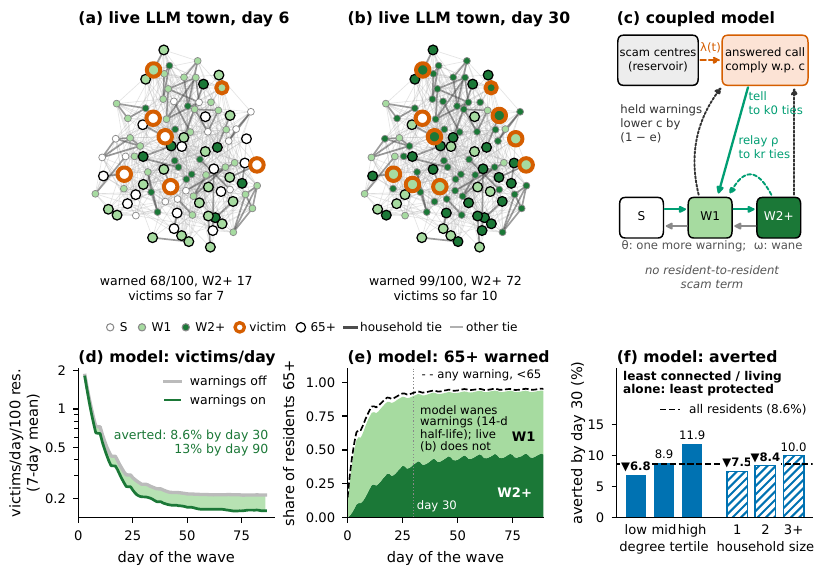}{3.9in}
\caption{\textbf{The town, the model and who is protected.} (a, b) One live LLM town (seed 705, P1, word of mouth on) on days 6 and 30. Fill shows how many different people have warned a resident about the scam they know most about (S none, W1 one, W2+ two or more); red rings mark victims; large nodes are residents aged 65+. (c) The two layers: scam calls, and tales told, relayed, believed and forgotten. (d--f) Fast simulator on the 20 consolidation-experiment town graphs at survey-level settings with 14-day forgetting, under the LLM-measured protection curve (scenario L, \S\ref{sec:model}): (d) victims per day, word of mouth off and on; (e) residents aged 65+ holding warnings (d and e run to 90 days); (f) victimisations averted by ties and household size: the least connected gain least.}
\label{fig:town}
\end{figure}

Scam Town has two coupled layers (Fig.~\ref{fig:town}c): scam calls, and the cautionary tales they provoke. Calls come from scam centres outside the town and never pass from resident to resident; in epidemiological terms the scam is a spillover threat with $\Rnot = 0$ \citep{lloydsmith2009epidemic}, so all of the spreading is done by the tales. We call a cautionary tale, once a resident believes and holds it, a \emph{warning}. A \emph{victimisation} is a complied scam call, and the \emph{attack rate} is victimisations per resident over a 30-day run. The same rules run in two simulators (Table~\ref{tab:split}). In \emph{live towns} an LLM plays every resident on every call. The \emph{fast simulator} replaces LLM call decisions by measured probabilities, to run thousands of towns. Appendix~\ref{app:model} gives the full model and Appendix~\ref{app:params} every parameter.

\textbf{The town} (Fig.~\ref{fig:town}a). A town has $N = 100$ synthetic residents, 45\% of them aged 65--95 (over-sampled; US share about 17\%). Residents live in households of one or two (older adults) or one to four (others) and are joined by five kinds of tie: household, family, neighbour, age-similar friend, and community ties through three hubs (a church, a senior centre and a pharmacy). A resident aged 65+ has 11.9 ties on average, against 15.9 for younger residents. These are \emph{possible} channels for a tale, not confidants (older Americans' core discussion network averages 3.6 people; \citealt{cornwell2008social}); the direction ``fewer, stronger ties'' is sourced \citep{wrzus2013social}, the levels assumed. Personas are drawn from random traits (technology literacy, shame, confiding, trust, loneliness).

\textbf{A simulated day.} A run lasts 30 days (ticks). Each day proceeds in five steps.
\begin{enumerate}[leftmargin=1.4em,itemsep=0pt,topsep=1pt]
\item \emph{Calls are scheduled.} Four scam centres each run one of ten documented scam patterns, with scripts written from 48 consumer-protection sources and grounded in analyses of real scam calls \citep{traister2026anatomy,traister2026corpus}. Each centre dials 4 residents per workday (16 scam calls per weekday; fewer at weekends) from a 40-name lead list. Lead lists over-weight residents aged 65+ and past victims, who are shared between centres (Table~\ref{tab:keyparams}). A centre switches or rewords its pattern when fewer than 8\% of its last 40 answered calls succeed. About 1--2 legitimate calls per resident per week run alongside.
\item \emph{The LLM answers} each call given the resident's persona and memories (below); victims go on the centres' ``sucker lists''. After a scam call it writes the resident's private takeaway, stored as a memory, and the message they would tell others.
\item \emph{Telling} (modelled). A resident tells anyone with probability 0.43 if a victim (survey evidence; \citealt{binette2026fraud}) and about 0.5 if they resisted, less with shame. In the live test each teller tells 2 people on average (3 in the what-if setting P2, \S\ref{sec:live}), chosen in proportion to tie preference and strength. An older victim who lost money avoids family.
\item \emph{Belief and memory} (modelled). Every warning from a personal tie or hub is believed, an upper bound (surveys suggest 0.60--0.85, a sensitivity); in live towns, warnings from institutions are believed with probability 0.5. A believed warning is stored as a memory in the sender's own words, and warnings about the same scam from different people merge into one memory (``2 people you know warned you about this''). Each memory carries a salience label (``vivid'', ``clear'', ``faded'') that decays with a 14-day half-life, and the 12 most salient enter the prompt; live-town memories are never deleted.
\item \emph{Relay} (modelled). Each believer passes the warning on, verbatim, the next day with probability $\rho = 0.2$ (0.4 at P2), to on average 2 (3) other ties, for at most six generations.
\end{enumerate}
A warning acts on a later call only through the LLM's prompt. In the fast simulator, seeding campaigns and broadcasts can also give chosen residents one warning before the run (\S\ref{sec:broadcast}).

\begin{table}[t]
\centering\small
\caption{\textbf{Who decides what.} LLM decisions are checked for plausibility (realism gates), not against human rates; two calibrations to real-world compliance failed. Social behaviour was modelled after the LLM failed at it (it believed 99.8\% of warnings; 82--93\% of LLM victims would tell someone), from the literature where possible, otherwise by assumption (Appendix~\ref{app:lessons}). The fast simulator replaces LLM call decisions by measured probabilities (Eq.~\ref{eq:comply}).}
\label{tab:split}
\begin{tabularx}{\linewidth}{@{}LL@{}}
\toprule
\textbf{LLM-driven} (Qwen3.7 Flash) & \textbf{Modelled probabilistically} (from literature or assumed) \\
\midrule
Answer to every scam and legitimate call: comply, verify, resist, hang up, or not answer & Town graph, households, tie strengths, personas (random traits) \\
After one's own scam call: the private takeaway and the wording of the warning & Call schedule, lead lists, 65+ targeting, sucker lists, pattern switching \\
& Whether a resident tells anyone, how many and whom \\
& Belief, passing on, maximum chain length \\
& Memory fading, merging and the prompt cap; education campaigns \\
\bottomrule
\end{tabularx}
\end{table}

\begin{table}[t]
\centering\small
\caption{\textbf{Key parameters} of the live test (P1, P2) and the fast simulator's survey-level settings. Only victims' telling is measured directly; the other levels are assumptions within literature bounds, converted to the model's units (Appendix~\ref{app:params}).}
\label{tab:keyparams}
\setlength{\tabcolsep}{3pt}
\begin{tabularx}{\linewidth}{@{}lllL@{}}
\toprule
Parameter & P1 (survey-level) / P2 (what-if) & Literature range & Source \\
\midrule
Telling, victims & 0.43 / 0.8 & 0.43 told family or friends & \citet{binette2026fraud,ianzito2025four} \\
Telling, resisters & formula, $\approx$0.5 / 0.8 & no direct measure & shape: \citet{parti2023if} \\
People told per teller & 2 / 3 & 1.6--3.6 close contacts & \citet{cornwell2008social} \\
Passing on $\rho$ & 0.2 / 0.4 & 0.05--0.35; $\leq 0.45$ & \citet{kumaraguru2009school,rader2012stories} \\
Belief, personal ties & 1.0 (upper bound) & 0.60--0.85 & \citet{shearer2024friends,silver2025americans} \\
Memory half-life & 14 days & 2 weeks to 4 months & \citet{scheibe2014forewarning,maertens2021long} \\
Scam calls & 4 centres $\times$ 4 per workday & no published rate & assumption \\
Targeting of 65+ & weight 2.5; past victims $\times 3$ & direction only & \citet{scheibe2014forewarning} \\
\bottomrule
\end{tabularx}
\end{table}

\textbf{The LLM resident.} The backbone is Qwen3.7 Flash (provider pinned, reasoning off). Each call is one request with a system frame (``you play one fictional resident\dots{} not as a fraud expert''), the persona in words, up to 12 memories, the caller ID and the script. The model returns JSON with an action (Table~\ref{tab:split}; ``verify'', hanging up to check through a known number, is offered only on calls that ask for something), a suspicion score and one sentence it would say (Appendix~\ref{app:prompts}).

\textbf{The fast simulator.} A CPU re-implementation of the same rules replaces each LLM call decision by a compliance probability:
\begin{equation}
c_i = c_{0,a}\, m_p\, (1 - \eown E_i)\,(1 - e_{D_i}),
\label{eq:comply}
\end{equation}
where $c_{0,a}$ is first-call compliance by age (0.344 for 65+, 0.065 otherwise), $m_p$ a pattern multiplier, $E_i$ an earlier answered call and \eown{} the protection it gives, and $e_D$ the protection from holding warnings from $D$ distinct senders. $c_{0,a}$ and \eown{} are measured in earlier live towns, $e_D$ in the LLM panel (\S\ref{sec:dose}). \textbf{Scenario L} (LLM-measured) uses $(e_1, e_2, e_3) = (0.27, 0.88, 0.96)$ with the in-town \eown{} of 0.92 (65+) and 0.98 (others): an LLM resident who has answered one scam call rarely falls for the next, unlike people. \textbf{Scenario H} (human-anchored) uses $(0.15, 0.40, 0.50)$ with $\eown = 0.15$; only $e_1$ comes from human trials, $e_2$ is an analogy from one trial of mailed warnings \citep{deliema2024fraud,langton2026calling}, and the rest is assumed (human-lower curve: $(0.15, 0.28, 0.39)$). The \emph{in-town curve}, measured inside earlier live towns, has a strong first warning ($e_1 \approx 0.63$ for residents aged 65+) and no extra second-dose effect (Appendix~\ref{app:panel}). Fast-simulator warnings are either \emph{forgotten} with a 14-day half-life or, as in live towns, kept; each result names its rule. The \emph{survey-level settings} are victims telling with probability 0.43 (surveyed), resisters about 0.5, two people told and $\rho = 0.2$ (assumed within literature bounds). Each warning yields on average
\begin{equation}
\Rw = \rho\, b\, \mathbb{E}[\text{people told}]
\label{eq:rwarn}
\end{equation}
further believed warnings, where $b$ is belief; talk sustains itself only if $\Rw > 1$.

\textbf{Measures and protocol.} The outcome is the relative reduction in attack rate against the same town without tales (for broadcasts, without any education). Most stages were declared with their predictions in a version-controlled log before their results were read; the exceptions, listed in Appendix~\ref{app:trail}, are the broadcast-belief amendment (A12.1, \S\ref{sec:broadcast}), the post-hoc batch inflation (A13), the live-test criteria committed after 51 of 80 towns had finished, and an undeclared comparison against the mean of both OFF arms. The fast simulator first passed a fidelity gate against 20 live towns at engine settings (Appendix~\ref{app:gate}). This is verification, not validation: the social layer is the same code, so the gate checks that the fast simulator reproduces the same LLM's behaviour, not that LLM residents behave like people.

\begin{figure}[t]
\centering
\mainfig{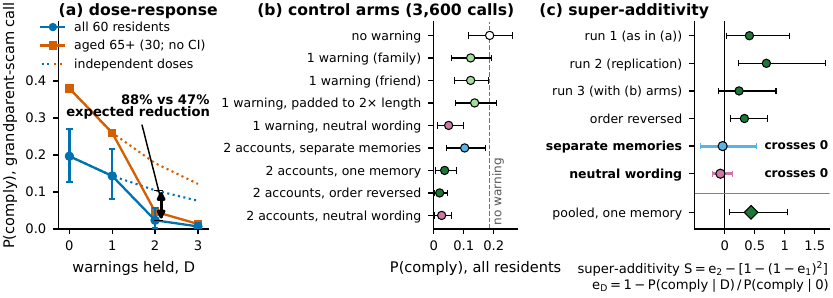}{2.0in}
\caption{\textbf{Exploratory dose-response of warnings in LLM residents (one scam pattern, one backbone).} (a) Compliance with a grandparent-scam call by warnings held, $D$ (60 residents, 5 calls per state; 95\% resident-cluster bootstrap; vermilion, the 30 aged 65+) against independent doses $1-(1-e_1)^D$: two warnings give $e_2 = 0.88$ against 0.47 expected. (b) Control arms (3,600 calls): padding one warning to the length of two did not lower compliance (0.137, as unpadded), two accounts in one memory did (0.037), and neutral wording makes one warning strong (0.050). (c) Super-additivity $S$ by run (run 1 is (a), run 2 a replication, run 3 the anchors of (b); diamond, pooled; intervals pass 1 where resamples give $e_1 < 0$). Pooled (two accounts, one standard-format memory) $S = 0.44$ [0.09, 1.05]; run 3 alone, separate memories and neutral wording cross zero.}
\label{fig:dose}
\end{figure}

\section{Results}

We report the results roughly in order of how directly they are grounded. The dose-response (\S\ref{sec:dose}) and the live test (\S\ref{sec:live}) are measured on LLM residents. The town-level baseline (\S\ref{sec:baseline}) and the phase maps (\S\ref{sec:phase}) are fast-simulator outputs at survey-level settings. The broadcast results (\S\ref{sec:broadcast}) are simulation only; broadcasts were never run live. ``Word of mouth alone'' takes several values across the paper because it depends on the memory rule, the protection curve and whether towns forget; Table~\ref{tab:womrecon} collects them in one place.

\subsection{How much does a cautionary tale protect?}
\label{sec:dose}

Sixty LLM residents each answered a grandparent-scam call five times in each of four memory states (Fig.~\ref{fig:dose}a). Compliance fell from 0.197 (95\% CI 0.127--0.270) with no warning to 0.143, 0.023 and 0.007 with one, two and three, almost all among residents aged 65+. One warning's effect was not distinguishable from zero ($e_1 = 0.27$, $-0.08$ to 0.55), but two protected far more than two independent doses would ($e_2 = 0.88$ against 0.47). Control arms (3,600 calls; Fig.~\ref{fig:dose}b) showed that this step came from a second \emph{account} of the scam merged into one memory, not a longer prompt, and need not come from a second person; held as two separate memories, it was not resolved. Wording also matters: written neutrally, one warning cut compliance to 0.050, and a second added no resolved benefit (Fig.~\ref{fig:dose}c). The super-additive shape is thus a property of the memory format, and a live consolidation test in 20 paired towns did not detect it (Appendix~\ref{app:friction}). These panels are exploratory: one pattern, one backbone, residents of one town, and several contrasts without adjustment. Super-additivity is resolved only when pooled ($S = 0.44$, 95\% CI 0.09--1.05); the single control-arm anchor ($S = 0.24$, $-0.09$ to 0.86) and the separate-memory arm both cross zero. LLM residents may differ from people, so later results carry both LLM and human curves.

\begin{figure}[t]
\centering
\mainfig{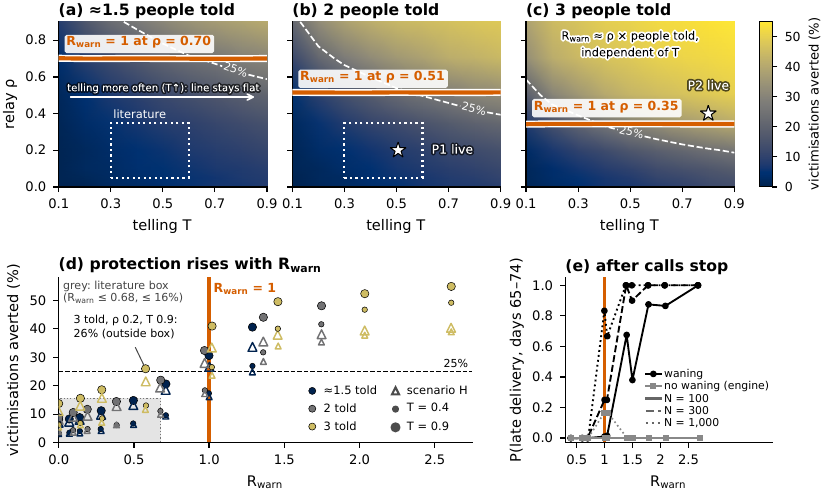}{3.4in}
\caption{\textbf{When does telling more give collective protection?} (a--c) Victimisations averted in a 30-day run with 14-day forgetting (scenario L, 20 graphs $\times$ 25 replicates per cell) over telling $T$ and relay $\rho$, for relayers telling about 1.5, 2 or 3 people. Vermilion: the critical line $\Rw = 1$ (simulated), at $\rho = 0.70$, 0.51 and 0.35; white dashed: 25\%; dotted box: literature range. Stars: positions of the live-test settings P1 and P2 only; their towns did not forget, so their reductions (21\%, 61\%) cannot be read off this map. (d) Each cell against its $\Rw$: literature-box cells have $\Rw \leq 0.68$ and avert at most 15.6\% (L) or 14.2\% (H); reductions saturate at 45--55\% (L) only at $\Rw \geq 2$. (e) If only deliveries from new senders are relayed, warnings persist after calls stop at day 15 (any believed delivery on days 65--74) only with waning memory, which recreates unwarned residents as in SIS models, at $\Rw \approx 1.4$ (towns of 100) or near 1 (larger towns).}
\label{fig:phase}
\end{figure}

\subsection{How much does word of mouth protect a town?}
\label{sec:baseline}
At survey-level settings with 14-day forgetting, word of mouth averted \textbf{8.6\%} of victimisations in a 30-day run under scenario L and \textbf{7.4\%} under scenario H (6.5\% with the human-lower curve). This is the paper's reference baseline. Universal belief inflates these figures; LLM residents' near-immunity after one scam call deflates them (14.4\% without that immunity; Appendix~\ref{app:robust}). Relay was weak ($\Rw = 0.38$), and the least-connected third of residents gained 6.8\% against 11.9\% for the best-connected (Fig.~\ref{fig:town}d--f). How much word of mouth protects depends mostly on memory (Table~\ref{tab:womrecon}): 7--9\% if warnings fade over two weeks, 12--15\% if they are kept (12.1\% here), and 19--21\% under the in-town curve and in live towns (Appendix~\ref{app:baseline}).

\begin{table}[t]
\centering\small
\caption{\textbf{Word of mouth alone: one reduction, read under different rules.} The same quantity (relative reduction in the 30-day attack rate from word of mouth, no broadcast) takes several values across the paper. The differences are configuration, not disagreement: they follow the memory rule, the protection curve and whether towns forget. Live towns do not forget.}
\label{tab:womrecon}
\begin{tabularx}{\linewidth}{@{}llllL@{}}
\toprule
Reduction & Memory & Curve & Forget? & Where \\
\midrule
8.6\% & 14-day half-life & L (survey-level) & yes & \S\ref{sec:baseline} reference; Appendix~\ref{app:baseline} \\
7.4\% & 14-day half-life & H & yes & \S\ref{sec:baseline}; Appendix~\ref{app:baseline} \\
12.1\% & kept (engine) & L & no & \S\ref{sec:baseline}; Appendix~\ref{app:baseline} \\
14.6\% & kept (engine) & panel curve, P1 graphs & no & Appendix~\ref{app:baseline}; forecast in \S\ref{sec:live} \\
19.1\% & kept (engine) & in-town table, P1 graphs & no & Appendix~\ref{app:baseline}; P1 forecast \\
21.0\% & kept (live towns) & in-town (live) & no & \S\ref{sec:live}, measured \\
\bottomrule
\end{tabularx}
\end{table}

\subsection{What would it take for education to protect strongly?}
\label{sec:phase}

Figure~\ref{fig:phase} maps protection over telling $T$, relay $\rho$ and the number of people told (498 settings, 14-day forgetting). The \emph{literature box} is $\rho$ 0.05--0.35, $T$ 0.3--0.6 and at most two people told \citep{kumaraguru2009school,rader2012stories}. No setting inside the box reached a 25\% reduction or had $\Rw$ above 0.68 (Fig.~\ref{fig:phase}d). Reaching 25\% took either relay up to the critical line $\Rw = 1$ ($\rho = 0.70$, 0.51 and 0.35 for about 1.5, 2 and 3 people told) or almost every callee telling three people (26\% at $T = 0.9$, $\rho = 0.2$). Tales are thus a weakly transmissible vaccine: they do not sustain themselves, but relay stretches each dose, so a seeding campaign needed about a quarter fewer seeds for the same effect (Fig.~\ref{fig:seed}a), as theory for weakly transmissible vaccines predicts \citep{basinski2019little}.

\subsection{Live test: word of mouth switched on and off in LLM towns}
\label{sec:live}

\begin{figure}[t]
\centering
\mainfig{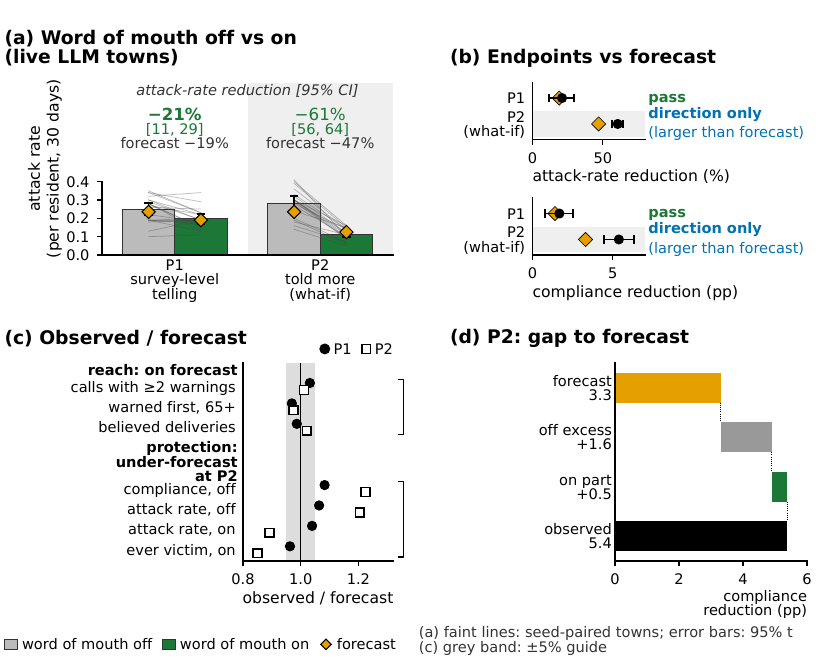}{4.4in}
\caption{\textbf{Live test: word of mouth switched off or on in 20 pairs of live LLM towns per setting (paired by seed; 30 days), against forecasts committed before any live result was read.} (a) Attack rate (victimisations per resident over 30 days) with word of mouth off (grey) and on (green); orange diamonds, forecasts. Survey-level settings (P1): 0.250 to 0.198, $-21$\% [11, 29], forecast 19\% ($[-3, 40]$ once batch-to-batch noise is included). What-if setting (P2, shaded; more telling and relay): 0.283 to 0.112, $-61$\% [56, 64], forecast 47\%. (b) Co-primary endpoints: P1 passes both; P2 is direction only (larger) on both, so the model is not forecast-validated. (c) Observed/forecast ratios (P1 filled, P2 open): reach within 0.97--1.03; protection under-forecast at P2. (d) Of the 2.1-point P2 compliance miss, 1.6 comes from the off arm.}
\label{fig:live}
\end{figure}

At each of two settings we ran 20 new seed-paired live LLM towns with word of mouth on and off. P1 uses the survey-level settings; P2 asks what if people told more (telling 0.8, three people told, $\rho = 0.4$; outside the literature box). Forecasts used the in-town curve; scenario L forecast 14.6\% at P1 and would also have passed, so the test checks the simulator, not the panel curve or the broadcast synergy. The forecasts were committed before any live result was read; the pass/fail criteria were committed after 51 of the 80 towns had finished, before any result was opened (Appendix~\ref{app:live}). A forecast \emph{passes} if its 95\% interval contains the predicted value, so a pass means ``the interval is consistent with the forecast'', not an equivalence test; the model is forecast-validated only if both co-primary endpoints pass at both settings. ``Not forecast-validated'' therefore means the forecasts were not met under this declared rule, not that word of mouth failed to protect.

Word of mouth protected the towns at both settings (Fig.~\ref{fig:live}a,b; Table~\ref{tab:ee}). At P1 the attack rate fell by 21.0\% (95\% CI 11.4--29.4\%; forecast 19\%) and compliance on answered calls by 1.7 percentage points (0.8--2.5; forecast 1.4): both endpoints pass. At P2 they fell by 60.6\% (56.4--64.4\%; forecast 47\%) and 5.4 points (forecast 3.3): both direction only, so the model is \textbf{not forecast-validated}. All four paired effects were significant ($p \leq 0.0004$, Holm), and reach was forecast almost exactly (Fig.~\ref{fig:live}c). Most of the P2 miss came from the word-of-mouth-off arm, which complied more than forecast (0.087 against 0.071; Fig.~\ref{fig:live}d). The two identically configured off arms differed by 1.0 point of compliance ($p = 0.003$), about 60\% of the P1 effect, so we also widened the P1 interval by this batch-to-batch variance: P1 then becomes $-1.7$ points of compliance [$-3.8$, $+0.5$] and a 21\% cut in attack rate [$-3$\%, 40\%]. Both include zero, so P1 cannot be separated from batch noise at 95\%, whereas P2 survives (61\% [49\%, 70\%]); P1 stays clearly positive only against the mean of both off arms (25.9\% [6.5\%, 41.3\%]; Appendix~\ref{app:live}).

\subsection{Broadcast education plus word of mouth}
\label{sec:broadcast}

\begin{figure}[t]
\centering
\includegraphics[width=0.85\linewidth]{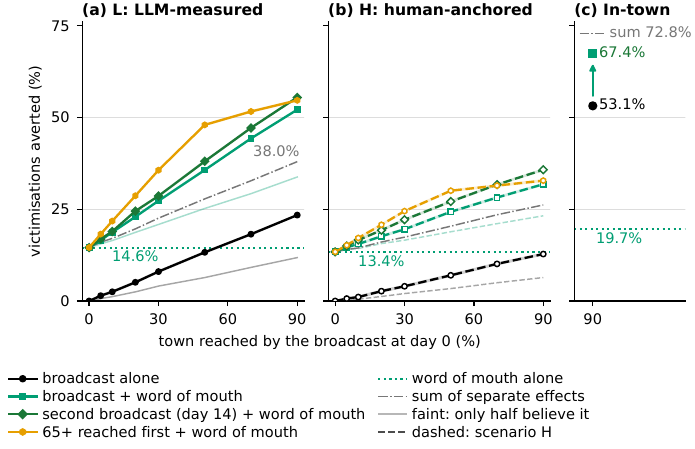}
\caption{\textbf{A broadcast averts more when people also talk about it} (fast simulator on the 20 live-test town graphs, no forgetting, 30 days, 60 replicates per town; Monte Carlo error about $\pm$0.5 points). Victimisations averted, against no education, when a trusted source warns a share of the town on day 0, believed by all (faint: by half). Black: broadcast alone; green: plus survey-level word of mouth (dark: a second broadcast on day 14; orange: residents aged 65+ reached first); dotted: word of mouth alone (14.6\% in a, 13.4\% in b); grey dash-dot: the sum of the two separate effects (38.0\% at 90\% in a); the gap above it is the synergy. (a) Scenario L; (b) scenario H (human-anchored): the same pattern, smaller. (c) In-town curve at 90\% coverage: 53.1\% alone, 67.4\% with word of mouth, below the 72.8\% sum: talk still adds 14 points, but the two overlap.}
\label{fig:broadcast}
\end{figure}

Unlike the live test in \S\ref{sec:live}, nothing in this subsection was run live. It extrapolates the live-checked fast simulator, and its headline synergy depends on the two-warning curve that the live consolidation test did not detect (\S\ref{sec:dose}). We read these as the most speculative results in the paper.

A simulated trusted broadcast warns a share of the town on day 0. Everyone reached believes it (an amendment after a first run at belief 0.5; Appendix~\ref{app:trail}), relays it like any warning when word of mouth is on, and, as in live towns, never forgets it. Reaching 90\% of the town, the broadcast alone averted 23.4\% (Fig.~\ref{fig:broadcast}a). With word of mouth it averted 52.1\%, more than the sum of the two (23.4\% + 14.6\% = 38.0\%): a synergy of $+14.2 \pm 0.7$ points. The synergy is set by the curve: a resident who hears the broadcast and a neighbour holds two accounts, which under L protect far more than one (\S\ref{sec:dose}). Under scenario H the pair averted 31.8\% against 12.8\% alone (synergy $+5.6$; Fig.~\ref{fig:broadcast}b). Under the in-town curve, where one warning already protects strongly, the broadcast alone averted 53.1\% and the pair 67.4\% (Fig.~\ref{fig:broadcast}c): the two overlap (synergy $-5.5$), though together they reach a 50\% cut at 52\% coverage. Talk thus added 14--29 points under every curve, and more than the sum only where a second warning protects far more than the first. Belief 1.0 is the most favourable case: at belief 0.5 the broadcast alone averts 11.8\% and the pair 33.8\%; with 14-day forgetting the pair averts 39.9\% (Appendix~\ref{app:robust}). Broadcasts were not tested live.

\textbf{Reach the people scammers call first.} With word of mouth, a broadcast to residents aged 65+ first needed 15\% coverage for a 25\% cut (random: 25\%). Seeding 20\% of residents added 12.8 points when they were those scammers would call (an oracle), 11.4 when aged 65+ and 4.6 when best-connected (Fig.~\ref{fig:seed}). Its size follows the modelled targeting and LLM compliance.

\section{Discussion and conclusion}
\label{sec:discussion}

It helps to separate what we measured from what we assumed. Measured on LLM residents: a cautionary tale protects its holder, and in live towns word of mouth cut victimisation by 21\% at survey-level settings (8.6\% once towns forget over 14 days) and by 61\% when people told more. Modelled from surveys and assumptions: who tells whom, who believes, and how warnings fade. Under those assumptions word of mouth never sustained itself ($\Rw \leq 0.68$), yet relay still stretched each dose. Simulation only, and never tested live: a trusted broadcast reaching 90\% of a town gained 14--29 points when people also talked about it, though that super-additive part reverses under the in-town curve. In this picture scam education behaves like a weakly transmissible vaccine.

\textbf{What this could mean for scam education.} If these patterns hold for people, three practical ideas follow, though each still needs a real-world trial. First, a warning should come with a story people can repeat, not just a tip: in our towns a second, concrete account of a scam protected far more than one longer warning. This was the one effect a live test did not confirm, so it is the least certain of the three. Second, campaigns should reach the people scammers actually call, and age is a practical guide to who that is; in our model this is almost true by definition, so it is really a reminder to target who gets called rather than who is most connected. Third, do not expect word of mouth on its own to protect the most isolated people, who gained the least in every version of the simulation. A real trial should also watch for a cost we saw: residents who had been warned also turned away about half of their legitimate calls (Appendix~\ref{app:lessons}), and that kind of over-caution could matter as much as the protection. None of these is yet evidence about people; each is a prediction a trial could confirm or overturn.

\textbf{Limitations.} The largest threat is stand-in validity. Our only test against held-out towns checks that the fast simulator reproduces the \emph{same LLM's} behaviour in full towns; it does not check that LLM residents behave like people, and no human outcome anchors any number here (Appendix~\ref{app:lessons}). The backbone is a single model, chosen non-blindly from five, and a system prompt that names fraud protection as the topic may cue protective answers. LLM call decisions are uncalibrated: victimisation is unrealistically high and was not varied, so relative effects hold for one scam rate, and residents become almost immune after one scam call (scenario H relaxes this). The dose-response uses one scam pattern, and every town is a static graph of about 100 residents. Resisters' telling, the number told and relay are assumed, belief sits at its upper bound, and the two-warning step is exploratory and undetected live. The broadcast headline uses belief 1.0, adopted after the declared run at 0.5, and its synergy reverses under the in-town curve. Monte Carlo intervals omit parameter uncertainty.

\section*{Ethics statement}
This study involved no human-subjects research; all participants are simulated LLM agents. Scam scripts were written from public consumer-protection descriptions, and the simulated centres only switch between or reword documented patterns; they do not optimise persuasion. Results come from LLM residents and are hypotheses for field trials, not evidence about how people respond to warnings.

\section*{Reproducibility statement}
Appendix~\ref{app:model} gives the full model, Appendix~\ref{app:params} every parameter with its source, Appendix~\ref{app:prompts} the prompts, model identifier and provider pin, Appendix~\ref{app:live} the live-test design, and Appendix~\ref{app:trail} the declaration trail with commit hashes. LLM responses are reproducible only up to provider sampling; the fast simulator is deterministic given its seeds.

\section*{AI use statement}
LLMs were the object of study, serving as simulated residents (Section~\ref{sec:model}). Generative AI tools also helped write analysis and figure code and draft text; all numbers were checked against the run outputs and the version-controlled results log, and the authors take responsibility for the content.

\section*{Data and Code Availability}
The datasets and code supporting this work are released by Scam.AI and are available from
the Scam.AI Research portal, \url{https://scam.ai/research}~\cite{scamai_research}. Document-
and PDF-forgery resources are additionally maintained at Trustable, \url{https://trustable.ai}~\cite{trustable_ai}.
These are the canonical, maintained locations for the resources associated with this paper and
supersede any prior ad hoc distribution.

\bibliography{references}
\bibliographystyle{IEEEtranN}

\clearpage
\appendix

\section*{Appendix contents}
\begin{enumerate}[label=\Alph*,leftmargin=2em,itemsep=0pt,topsep=2pt]
\item Full model and equations \hfill \ref{app:model}
\item Parameters and sources \hfill \ref{app:params}
\item Fidelity gate (verification) \hfill \ref{app:gate}
\item Baseline and persistence \hfill \ref{app:baseline}
\item Robustness \hfill \ref{app:robust}
\item Scammer economics (extension) \hfill \ref{app:econ}
\item Dose-response panels and control arms \hfill \ref{app:panel}
\item Live test: design and full results \hfill \ref{app:live}
\item Friction and the consolidation test \hfill \ref{app:friction}
\item Declaration and amendment trail \hfill \ref{app:trail}
\item What we tried, and LLM methods lessons \hfill \ref{app:lessons}
\item LLM requests and prompts \hfill \ref{app:prompts}
\item Compute and cost \hfill \ref{app:cost}
\end{enumerate}

\section{Full model and equations}
\label{app:model}

\begin{figure}[H]
\centering
\includegraphics[width=\linewidth]{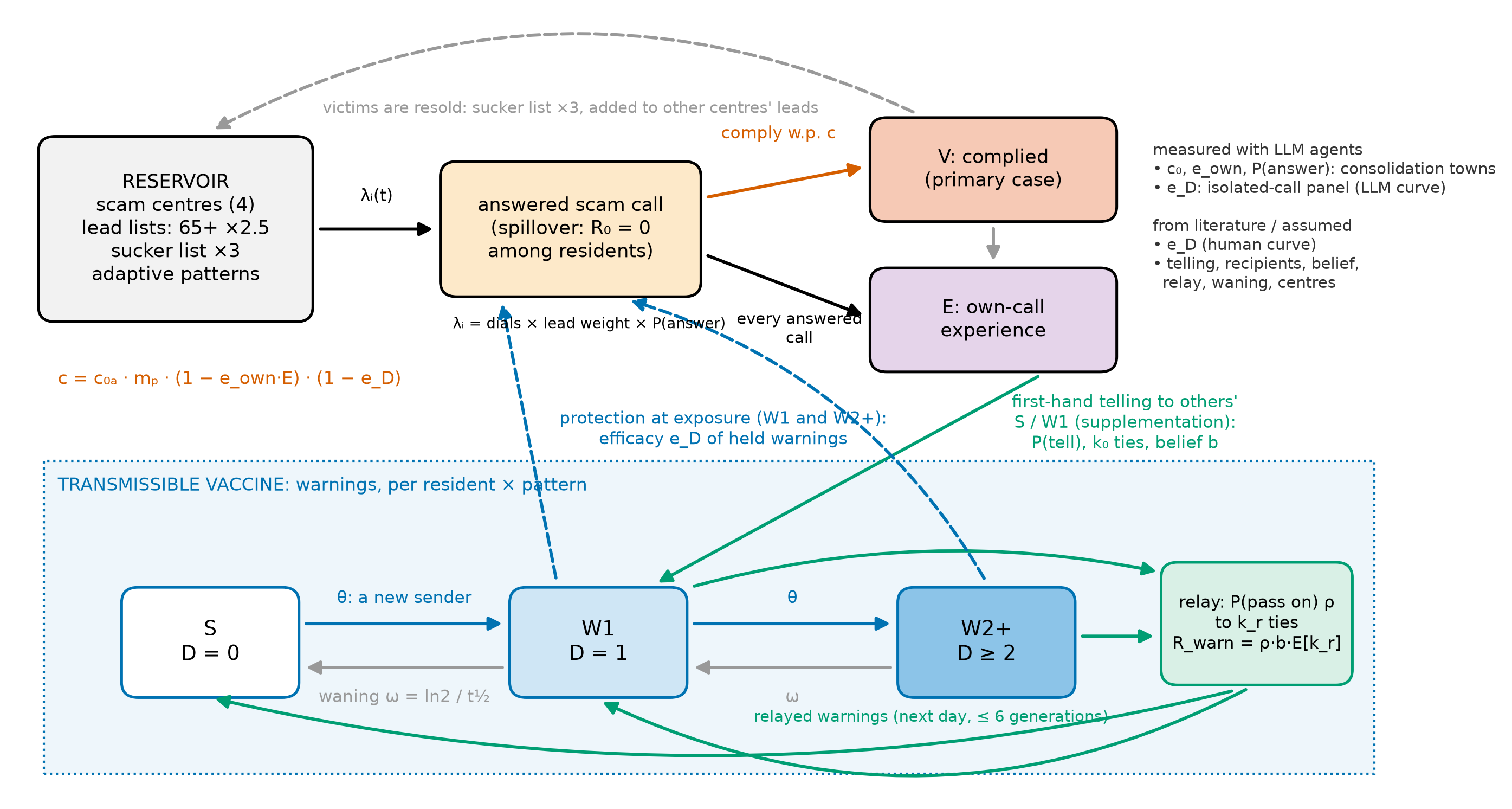}
\caption{\textbf{Full model structure.} Scam centres (the reservoir) dial residents from lead lists that over-weight residents aged 65+ and resell victims (sucker list). Each answered call is a spillover event. The callee complies (becomes V) with probability $c$, which falls with their own earlier answered calls (E) and with the number $D$ of distinct warnings they hold about the call's pattern. Answered calls may generate first-hand warnings to the callee's ties, moving recipients from S to W1 and W2+. Believers relay with probability $\rho$, and held warnings wane at rate $\omega$. Protection at exposure acts from W1 and W2+.}
\label{fig:schematic}
\end{figure}

\paragraph{Notation.} W denotes warned (vaccinated) residents and V complied calls (victims). In the transmissible-vaccine literature V denotes vaccine-infected hosts \citep{nuismer2016eradicating,basinski2019little}; we keep V because it labels complied calls in every figure. Compartments are tracked per resident and per scam pattern $p$: S ($D = 0$), W1 ($D = 1$) and W2+ ($D \geq 2$). Here $D$ is the number of distinct believed senders whose warning about $p$ the resident still holds. E marks residents who have answered at least one earlier scam call of any pattern. $T$ is the probability of telling, $H$ the length of a wave, $\gamma(t)$ the rate of first-hand warnings, $\bar k$ a set mean number of people told, $m$ the mean held depth, $\rho$ relay and \rhopop{} call-weighted protection.

\paragraph{Reservoir.} At $N = 100$, each centre places four calls per workday (80\% of them Monday to Friday) from a 40-lead list. Leads are drawn with weight $1 + 1.5\cdot[\text{age} \geq 65]$. At $N = 300$ and $1{,}000$, dials and leads scale with $N$. Victims are added to the sucker list with weight $\times 3$, and to each other centre's leads with probability 0.5. A lead that fails to answer three times is replaced. A centre switches or paraphrases its pattern when its hit rate over 40 answered calls falls below 0.08. The force of infection on resident $i$ in age group $a$ is
\begin{equation}
\lambda_i(t) = d(t)\, \Pr(i \text{ drawn from the leads})\, \Pr(\text{answer} \mid a, E_i),
\label{eq:foi}
\end{equation}
the product of reservoir pressure, targeting and exposure \citep{lloydsmith2009epidemic,plowright2017pathways}. Eleven kinds of legitimate call run alongside the scam wave in the live towns, so that friction can be measured.

\paragraph{Fast-simulator compliance, fully specified.}
Each call is first answered or not, and, if answered, complied with or not. A resident of age class $a$ (65+ or under 65) who has answered an earlier scam call ($E_i = 1$) answers with probability $\Pr(\text{answer}\mid a, E_i)$, a constant for each of the four $(a, E_i)$ cells and nothing else: 0.969 (65+, $E_i = 0$), 0.925 (65+, $E_i = 1$), 0.987 (under 65, $E_i = 0$) and 0.921 (under 65, $E_i = 1$), each the observed share of scam calls answered in that cell of the live no-education towns. An answered call is complied with with probability $c_i$ of Eq.~\ref{eq:comply}, clipped to $[0,1]$. The first-call compliance $c_{0,a}$ is 0.344 for 65+ and 0.065 for under 65, the compliance on answered calls of residents with $D = 0$ and $E_i = 0$. The pattern multiplier $m_p$ is a fitted constant for each of the ten scam patterns, set by iterative proportional fitting to the same answered calls and shrunk toward 1 by a prior worth 20 calls at the mean compliance rate (a pattern absent from the data gets $m_p = 1$); the fitted values are $m_p = 0.77, 1.28, 1.79, 0.83, 0.36, 1.07, 0.27, 0.78, 0.82, 1.41$ for patterns P01--P10. $E_i$ is a binary indicator: it becomes 1 after the resident's first answered scam call of any pattern and takes effect from the next simulated day, so the factor $(1 - \eown E_i)$ is applied at most once and never compounds with the number of earlier calls; $\eown = 0.92$ for 65+ and 0.98 for under 65. $D_i$ is the number of distinct believed senders whose warning about the called pattern the resident currently holds, and $e_{D_i}$ is read from the scenario's curve $(e_1, e_2, e_3)$ with $e_0 = 0$. The curve saturates: every $D \geq 3$ uses $e_3$, with no cap and no extrapolation, so in scenario L a resident holding four or more warnings keeps a fraction $1 - e_3 = 0.036$ of the otherwise applicable compliance. The in-town curve is the exception, with $e_1 \approx 0.63$ for residents aged 65+ and no extra second-dose effect ($e_{D} = e_2$ for all $D \geq 2$; Appendix~\ref{app:panel}). The fidelity gate of Appendix~\ref{app:gate} instead replaces the product by an in-town table of $c_i$ indexed by age, $E_i$ and $\min(D, 2)$, multiplied by $m_p$, with $D$ counted over all warnings ever held. The force of infection of Eq.~\ref{eq:foi} is the number of dials per day, times the probability that resident $i$ is drawn from a centre's lead list, times $\Pr(\text{answer}\mid a, E_i)$ above. There is no resident-to-resident scam transmission.

\begin{figure}[H]
\centering
\includegraphics[width=\linewidth]{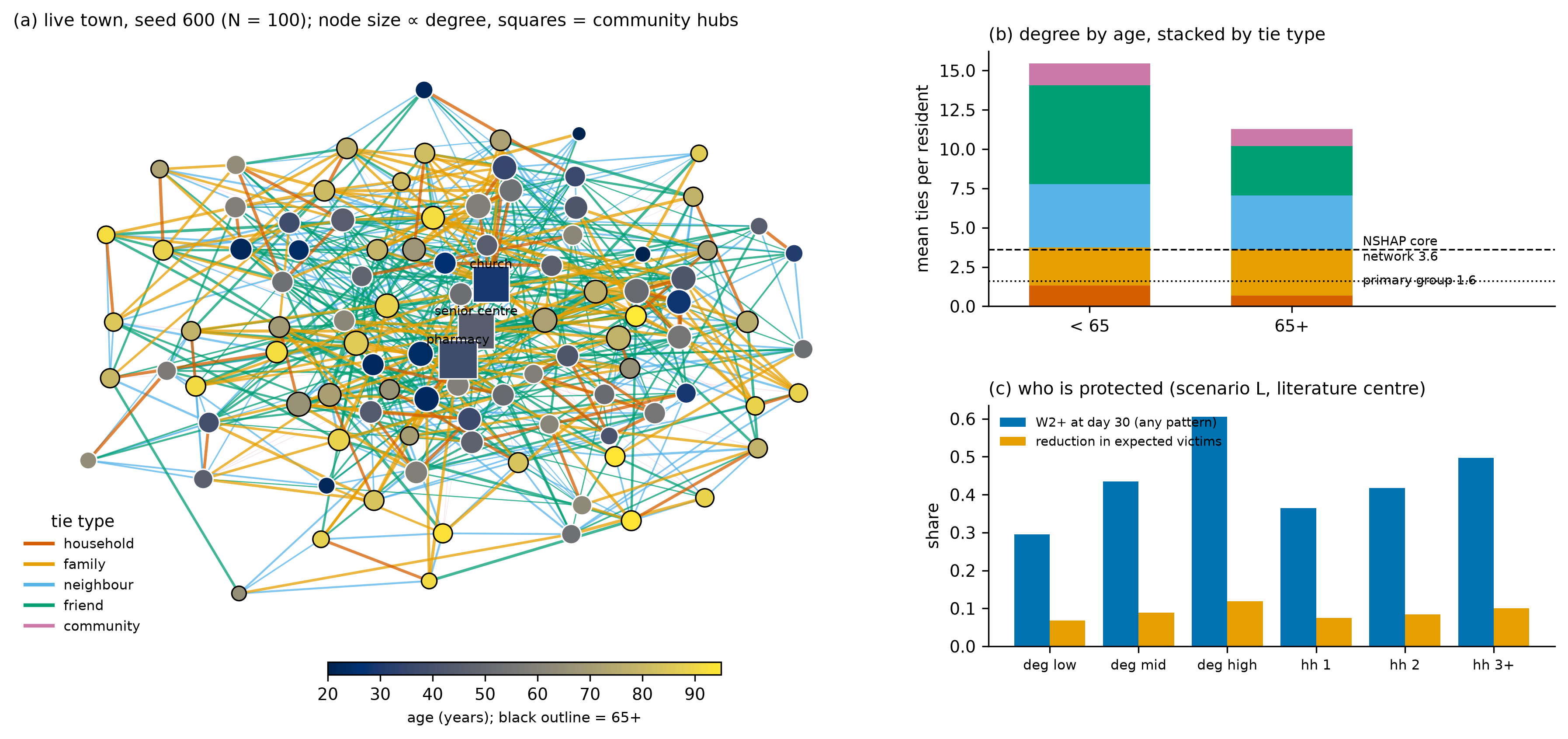}
\caption{\textbf{The town.} (a) One live town (seed 600); node colour is age, squares are community hubs, and edge colour is tie type. (b) Mean ties per resident by age group and tie type, against the core discussion network (3.6) and primary group (1.6) of older adults \citep{cornwell2008social}. (c) Share in W2+ at day 30 and reduction in expected victimisations, by degree tertile and household size.}
\label{fig:towngraph}
\end{figure}

\paragraph{Town.} Older households have one resident with probability 0.45 and otherwise two (29\% of older residents live alone on the consolidation-experiment graphs); other households have one to four. Household ties link everyone in a home, and family ties link each older household to one or two younger ``adult-child'' households. Neighbours sit on a shuffled ring, each household linked to its two (older) or three (others) nearest households. Friends are age-homophilous, about 1.5 per older resident and 3.5 per other resident. Three community hubs (church, senior centre, pharmacy), held by the best-connected adults, connect their members: 45\% of older residents and 15\% of others. Mean degree is 11.9 for residents aged 65+ and 15.9 for others on the consolidation-experiment graphs (11.7 and 15.8 on the live-test graphs), with kin shares of 0.34 and 0.27 (Fig.~\ref{fig:towngraph}). Tie strengths are in Appendix~\ref{app:params}. The tie counts and strengths are assumptions; their directions follow the network literature \citep{cornwell2008social,wrzus2013social}.

\paragraph{Generation, belief, relay and memory.} After each answered call the callee tells others with probability $P(\text{tell})$. The default is a calibrated formula (Appendix~\ref{app:params}) giving about 0.14--0.17 for older victims and 0.46--0.55 for older resisters. The literature set raises victims to 0.43 \citep{binette2026fraud,ianzito2025four}. First-hand recipients number $k_0 = 1 + \mathrm{Bin}(3, 0.5q)$ under the engine rule, where $q$ is disclosure propensity, or $k = 1 + \mathrm{Bin}(4, (\bar k - 1)/4)$ for a set mean $\bar k$. Hubs tell two more. First-hand warnings per resident per day are
\begin{equation}
\gamma(t) = \frac{1}{N}\sum_i \lambda_i(t)\,\bigl[c_i\, p^{V}_{i} + (1 - c_i)\, p^{R}_{i}\bigr]\, \mathbb{E}[k_0]\, b,
\label{eq:gen}
\end{equation}
where $p^V_i$ and $p^R_i$ are the telling probabilities of resident $i$ as a victim and as a resister. Because victims are 1--12\% of tellers, $\gamma$ is set mainly by resisters. A believer relays the received text verbatim, with probability $\rho$, on the next day, to $k_r$ ties other than the sender, for at most six relay generations. A repeat warning from the same sender does not add depth. In the main model every believed delivery is relayed with probability $\rho$, so the relay layer is a branching process on messages. The \emph{engine memory rule} consolidates warnings about one pattern into one memory, and nothing is dropped within 50 days. The \emph{per-warning rule} lets each held warning lapse at rate $\omega = \ln 2/t_{1/2}$, with $t_{1/2} = 14$ days. Human evidence spans single light doses that fade within weeks to intensive doses with half-lives of six weeks to four months \citep{scheibe2014forewarning,maertens2021long}.

\paragraph{Heuristics.} With distinct-sender arrivals at rate $\theta$ and per-warning waning, depth follows an infinite-server chain,
\begin{equation}
\frac{dX_D}{dt} = \theta X_{D-1} - \theta X_D - D\omega X_D + (D+1)\omega X_{D+1},
\label{eq:depth}
\end{equation}
whose stationary state is Poisson with mean $m = \theta/\omega$. Victim incidence is $dV_a/dt = \sum_{i \in a} \lambda_i(t)\, c_i(t)$, and the attack rate over a wave of length $H$ is $\mathrm{AR} = N^{-1}\int_0^{H} \sum_i \lambda_i c_i\, dt$, so repeat victimisation counts. Below $\Rw = 1$, each answered call produces on average
\begin{equation}
\mathbb{E}[\text{believed warnings per answered call}] = \frac{P(\text{tell})\,\mathbb{E}[k_0]\, b}{1 - \Rw}
\label{eq:cascade}
\end{equation}
believed warnings. The simulation matched this to three decimals (1.256 against 1.257 at the engine defaults; 1.679 against 1.684 at the literature centre). As a check, $\Rw$ was also computed as the dominant eigenvalue of a 10-type next-generation matrix (age $\times$ arrival tie type) \citep{diekmann1990definition}. This next-generation estimate $R_{\mathrm{ngm}}$ agrees with the branching estimate $R_{\mathrm{branch}}$ from Eq.~\eqref{eq:rwarn} to within 0.001 in every literature-box cell and to within 0.0016 across all 498 cells (maximum relative difference 0.8\%, within Monte Carlo error), so the conclusion that no literature-box cell is self-sustaining (largest $\Rw$ 0.68) does not depend on the estimator. At quasi-steady state, mean depth is about $m \approx \gamma\tau/(1 - \Rw)$ with $\tau = 1/\omega$, the closed form of ``sharing outpaces decay'' \citep{pilditch2022psychological}. Under Poisson depth, $\rhopop(m) \approx e_1 m + (e_2 - 2e_1)m^2/2$ for small $m$, which is convex when $e_2 > 2e_1$. Simulated depth was over-dispersed (\phitwo{} was 1.2--1.3 times its Poisson value), so these heuristics are guides, not results.

\paragraph{Measures.} Protective coverage at exposure, \phitwo, is the share of dialled calls that reach a W2+ resident. Call-weighted protection is $\rhopop = 1 - \bar\pi/\bar\pi_0$, the relative reduction in compliance on answered calls against a town without warnings. Warning beats the scam (WBS) is the share of (resident, pattern) pairs with at least one answered call of the pattern in which the first believed warning arrived strictly before that call \citep{rozema2025anti,nuismer2018controlling}. The \emph{direct} part of the reduction in attack rate sums $c(D = 0) - c(D)$ over answered calls on the warning path, at the same own-call state; the \emph{indirect} part is the remainder.

\section{Parameters and sources}
\label{app:params}

\begin{table}[H]
\centering\scriptsize
\caption{\textbf{Parameters, values and sources.} [S] sourced; [A] assumption; [A, dir S] direction sourced, level assumed. ``Engine'' is the configuration of the live consolidation towns; ``lit'' is the literature centre.}
\label{tab:params}
\begin{tabularx}{\linewidth}{@{}p{2.4cm}p{2.7cm}Lp{1.3cm}@{}}
\toprule
Parameter & Value (engine; lit; sweep) & Evidence & Status \\
\midrule
Victim $P(\text{tell})$ & formula $\approx$0.14--0.17 (65+); 0.43; 0.1--0.9 & 43\% of victims warned family or friends (survey releases) \citep{binette2026fraud,ianzito2025four}; 77--84\% told someone \citep{morgan2021financial,nij2024examining}; older victims who lost money describe concealment \citep{parti2023if} & [S] level \\
Resister $P(\text{tell})$ & formula $\approx$0.46--0.55; kept; global $T$ 0.1--0.9 & no direct measure; 73\% discuss local crime with friends, family and neighbours \citep{shearer2024friends} & [A], decisive \\
First-hand recipients $k_0$ & mean 1.8; 2; \{1.5, 2, 3\} & core discussion network 3.6 and primary group 1.6 in older adults \citep{cornwell2008social}; 2.08 core confidants for US adults \citep{mcpherson2006social} & [A, bounded S] \\
Relay recipients $k_r$ & mean $\approx$1.46; 2; \{1.5, 2, 3\} & none & [A], decisive for $\Rw = 1$ \\
Belief $b$ & 1.0; 1.0; 0.85/0.75/0.60 & 72\% judge friends' and family's crime news accurate \citep{shearer2024friends}; 44\% trust most neighbours \citep{silver2025americans} & [A] upper bound \\
Relay $\rho$ & 0.2; 0.2; 0--0.9 & converted: 0.13 discussed tips within 30 days \citep{kumaraguru2009school}; $\leq$0.45 retold a recalled story \citep{rader2012stories}; per-neighbour passing 0.05--0.35 is a different unit \citep{banerjee2013diffusion} & [S] band \\
Max relay generations & 6 (removed in the persistence test) & none & [A] \\
Half-life of one warning & none within 50 d (engine); 14 d; 7--45 d & light doses fade within weeks \citep{scheibe2014forewarning,robb2023who}; intensive doses about 6 weeks to 4 months \citep{maertens2021long,reinheimer2020investigation} & [S] range \\
Half-life after 2nd warning & as one; 30--60 d sensitivity & boosted residual effects \citep{burke2022can,maertens2021long} & [A, dir S] \\
Efficacy $e_1, e_2, e_3$ & LLM 0.27/0.88/0.96; human 0.15/0.40/0.50; human-lower 0.15/0.28/0.39 & isolated-call panel (P02, one backbone); human $e_1$ \citep{deliema2024fraud,akesson2023preventing}; $e_2$ by analogy from six contacts against one in one USPIS trial (abstract \citep{deliema2024fraud}; journal paper of the same trial \citep{langton2026calling}) & LLM-measured; [A] ratio \\
Own-call efficacy $\eown$ & in-town 0.92 (65+)/0.98; 0.15 in scenario H & consolidation-experiment logs; repeat victimisation in humans & LLM-measured; [A] \\
First-call compliance $c_{0,a}$ & 0.344 (65+), 0.065 & cross-fitted in-town table & LLM-measured \\
Warning fatigue & off; $f \in \{0.1, 0.25\}$ & identical warnings habituate \citep{akesson2023preventing,vance2018tuning} & [S] direction \\
Cross-pattern spillover & 0 & different-scam warnings partly transfer \citep{scheibe2014forewarning,banas2010meta} & [S] direction \\
Scam pressure & 4 centres $\times$ 4 dials/workday, 40 leads, 65+ $\times$2.5, sucker $\times$3; $\times\{0.5, 2\}$ & no published volume per older adult found & [A] \\
Break-even $R_{\mathrm{scam}}$ (extension) & \{1.5, 2, 3, 8\} & none & [A] \\
Share aged 65+ & 0.45; sensitivity 0.20 & US $\approx$0.17; deliberate over-sample & [A] design \\
Pattern multiplier $m_p$ & P01--P10: 0.77, 1.28, 1.79, 0.83, 0.36, 1.07, 0.27, 0.78, 0.82, 1.41; unseen pattern 1 & fitted to answered calls of the 20 live no-education towns, shrunk to 1 with a prior of 20 calls & LLM-measured \\
$\Pr(\text{answer}\mid a, E)$ & 65+: 0.969 ($E{=}0$), 0.925 ($E{=}1$); under 65: 0.987, 0.921 & share of scam calls answered in each cell of the same towns & LLM-measured \\
$e_D$ for $D \geq 4$ & $e_3$ (saturates; no cap or extrapolation); in-town $e_D = e_2$ for $D \geq 2$ & model rule, not measured at $D \geq 4$ & [A] \\
\bottomrule
\end{tabularx}
\end{table}

\paragraph{Disclosure formula and engine details.} The default telling probability is
\[\small
P(\text{tell}) = \sigma\bigl(-0.2 + 1.6(q - 0.5) + 0.6\,[\text{resisted, no loss}] - 1.5\, s\, (1 \text{ if loss else } 0.4) - 0.6\, s_{65}\,[65{+},\ \text{loss}]\bigr),
\]
where $q$ is disclosure propensity and $s$ shame (shape from \citealt{parti2023if}; coefficients assumed). Persona traits are Beta draws with concentration 6. Means are: technology literacy 0.35 for residents aged 65+ and 0.65 for others; shame 0.6 and 0.4; disclosure propensity 0.4 and 0.6. Tie strengths are drawn uniformly:

\begin{center}\small
\begin{tabular}{@{}ll@{}}
\toprule
Tie & Strength \\
\midrule
household & U(0.85, 1.0) \\
family & U(0.70, 0.95) \\
neighbour & U(0.45, 0.70) for 65+; U(0.30, 0.55) otherwise \\
friend & U(0.6, 0.9) and U(0.3, 0.7) \\
community & U(0.30, 0.55) \\
\bottomrule
\end{tabular}
\end{center}

Recipient weights are household 3, family 2, friend 1.5, neighbour 1 and community 1. The persona's preferred confidant gets $\times 2$, and family gets $\times 0.3$ for an older victim who lost money. Memory strengths of social warnings are family 0.6, friend 0.5, and neighbour and hub 0.4. A prompt shows at most 12 memories.

The \emph{literature centre} sets: victim telling 0.43, the formula for resisters (mean about 0.5), two people told for both first-hand telling and relay, $\rho = 0.2$, and per-warning waning with $t_{1/2} = 14$ days. The literature box used in the parameter maps is $\rho \in [0.05, 0.35]$, $\bar k \leq 2$ and $T \in [0.3, 0.6]$. A sensitivity extends it to $\rho = 0.5$.

\section{Fidelity gate (verification)}
\label{app:gate}
Before any experiment, the closed-loop model had to reproduce the 20 live consolidation-experiment towns at their own settings. These were engine-default parameters, engine-rule memory, cross-fitted in-town tables, and adaptive centres as run, with 200 replicates per town. The declared gate required all four parts (Table~\ref{tab:gate}).

G-A1 and G-A2 test that the re-implementation, with calls generated by the engine's centre rules, reproduces live propagation given the LLM's call outcomes; nothing was fitted to depth. G-A3 and G-A4 are largely circular, because the compliance tables come from the same towns, out of fold. Per town, simulated and observed \phitwo{} correlated at $r = 0.70$, and per-town compliance at $r = 0.46$. Dials per town matched (357 against 357), as did believed deliveries (420 against 430). The simulated per-town ON$-$OFF difference had a standard deviation of 0.005, against 0.019 in the live pairs, which carry LLM sampling noise. Called residents were no more often warned than random residents (overlap ratio 0.97).

\begin{table}[H]
\centering\small
\caption{\textbf{Fidelity gate.} Closed-loop model on the 20 live consolidation-experiment graphs (200 replicates per town per arm) against the observed live towns. All four declared gates passed.}
\label{tab:gate}
\resizebox{\linewidth}{!}{\begin{tabular}{@{}lllll@{}}
\toprule
Gate & Simulated & Observed (live) & Tolerance & Verdict \\
\midrule
G-A1 depth 0 / 1 / 2 / 3+ (dialled calls) & 0.668 / 0.232 / 0.073 / 0.027 & 0.655 / 0.240 / 0.077 / 0.027 & $\pm$0.03 each & pass \\
G-A2 \phitwo & 0.100 & 0.104 & $\pm$0.02 & pass \\
G-A3 compliance, answered calls & 0.0617 & 0.0615 & $\pm$0.01 & pass \\
G-A4 consolidation ON $-$ OFF & $-0.0060$ & $-0.0057$ [$-0.0145$, $+0.0031$] & inside CI & pass \\
\bottomrule
\end{tabular}}
\end{table}

\begin{figure}[H]
\centering
\includegraphics[width=\linewidth]{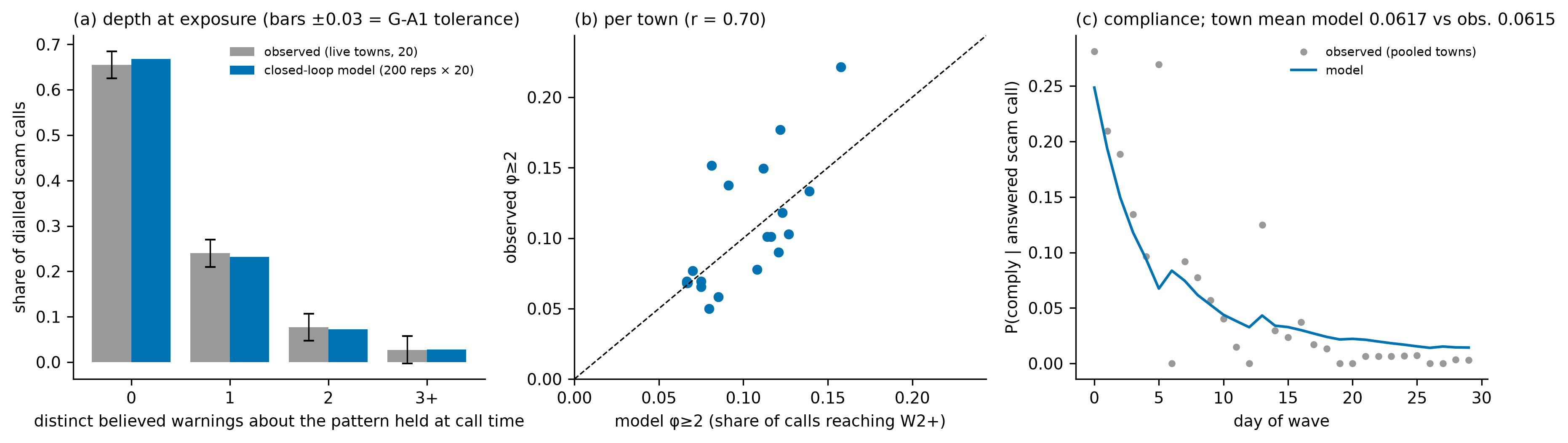}
\caption{\textbf{Fidelity gate.} (a) Depth distribution of dialled scam calls in the 20 live towns (grey; bars show the $\pm$0.03 gate tolerance) and in the model (blue). (b) Per-town share of calls reaching a twice-warned resident, model against observation ($r = 0.70$). (c) Compliance on answered calls by day of the wave.}
\label{fig:gate}
\end{figure}

\section{Baseline and persistence}
\label{app:baseline}

\begin{table}[H]
\centering\small
\caption{\textbf{Baseline collective protection.} Relative reduction in the attack rate (AR) against a town without warnings, on 20 live graphs (100 replicates per graph at 30 days, 50 at 90 days). ``Towns'' gives the min / median / max of per-town reductions. The 30-day literature-centre rows are the paper's reference values. Relative reductions carry Monte-Carlo error of about $\pm$0.005, so the literature-centre scenario-L value appears as 0.082--0.086 across runs (it is quoted as 0.086 in the main text, 0.085 in Appendix~\ref{app:robust}, and 0.082 in the memory-rule contrast).}
\label{tab:baseline}
\resizebox{\linewidth}{!}{\begin{tabular}{@{}llllllll@{}}
\toprule
Scenario & Days & Parameters & AR, no warnings & Relative reduction & Towns & \phitwo & $\Rw$ / WBS \\
\midrule
L & 30 & engine defaults & 0.237 & 0.088 & 0.06 / 0.08 / 0.13 & 0.100 & 0.28 / 0.23 \\
L & 30 & \textbf{literature centre} & 0.237 & \textbf{0.086} & 0.06 / 0.08 / 0.12 & 0.082 & 0.38 / 0.28 \\
L & 90 & engine defaults & 0.369 & 0.199 & 0.15 / 0.20 / 0.24 & 0.264 & 0.29 / 0.33 \\
L & 90 & literature centre & 0.369 & 0.134 & 0.09 / 0.13 / 0.20 & 0.133 & 0.40 / 0.39 \\
H & 30 & engine defaults & 0.768 & 0.080 & 0.05 / 0.07 / 0.12 & 0.111 & 0.28 / 0.25 \\
H & 30 & \textbf{literature centre} & 0.768 & \textbf{0.074} & 0.04 / 0.07 / 0.12 & 0.093 & 0.39 / 0.31 \\
H & 90 & engine defaults & 2.40 & 0.187 & 0.09 / 0.17 / 0.27 & 0.348 & 0.29 / 0.35 \\
H & 90 & literature centre & 2.40 & 0.117 & 0.04 / 0.11 / 0.18 & 0.185 & 0.40 / 0.43 \\
\bottomrule
\end{tabular}}
\end{table}

\paragraph{Uncertainty, absolute rates and the direct/indirect split.}
\begin{itemize}[leftmargin=1.2em,itemsep=1pt,topsep=2pt]
\item \emph{Panel uncertainty.} Propagating it through 200 bootstrap draws of the curve gave a 95\% uncertainty interval of 0.028--0.132 at the literature centre (median 0.096), 0.008--0.120 at the engine defaults and 0.216--0.351 near threshold. Uncertainty in the in-town tables was not propagated.
\item \emph{Absolute rates.} Absolute attack rates in scenario H (0.77 per resident per 30 days) are not realistic. Rescaling first-call compliance to human-plausible levels (0.01--0.165 for residents aged 65+) left the relative reduction at 0.064--0.079.
\item \emph{Direct and indirect effects.} At 30 days in scenario L, the reduction in attack rate was 0.0205 victimisations per resident, of which 0.0010 (5\%) was indirect. Over 90 days in scenario H the indirect part became negative, about $-10\%$: residents protected by warnings stayed off sucker lists, and calls were redirected to others.
\item \emph{Timing of protection.} In scenario L, 59--62\% of averted victimisations occurred on a resident's first answered call, and 41--49\% involved a relayed warning.
\item \emph{Town size and age mix.} In generated towns of 100, 300 and 1,000 residents, the literature-centre reduction was 8.5\%, 7.4\% and 6.7\% in scenario L and 7.6\%, 8.0\% and 9.4\% in scenario H. With 20\% of residents aged 65+ it rose to 0.117 (L) and 0.102 (H).
\item \emph{Memory rule.} Switching the literature centre to the engine memory rule raised the reduction from 0.082 to 0.121 in scenario L and from 0.075 to 0.111 in scenario H.
\end{itemize}

\paragraph{Why the live P1 forecast (19\%) exceeds the headline (8.6\%).} P1 uses the engine memory rule and the in-town table of the consolidation-ON arm, whose first warning is strong ($e_1 \approx 0.63$ among first calls to residents aged 65+). On the P1 towns, the panel curve gave 14.6\% and the in-town table 19.1\%. The gap between the two numbers is therefore a difference of configuration, not a discrepancy.

\begin{table}[H]
\centering\small
\caption{\textbf{Where $\Rw = 1$ sits} (scenario L, 20 graphs $\times$ 25 replicates; ``simulated'' is the measured branching ratio). $\Rw$ depends only on relay and the number told, not on telling.}
\label{tab:contour}
\begin{tabular}{@{}llll@{}}
\toprule
Mean number told & $\Rw$ at $\rho = 0.2$ & $\rho$ at $\Rw = 1$ (simulated) & Predicted before the run \\
\midrule
engine rule ($\mathbb{E}[k_{r,\mathrm{eff}}] \approx 1.43$) & 0.28 & 0.70 ($\Rw = 1.00$) & $\approx$0.7 \\
2 & 0.39 & $\approx$0.51 (0.97 at $\rho = 0.5$) & 0.5--0.55 \\
3 & 0.58 & $\approx$0.35 (1.02 at $\rho = 0.35$) & 0.35--0.4 \\
\bottomrule
\end{tabular}
\end{table}

\begin{figure}[H]
\centering
\includegraphics[width=\linewidth]{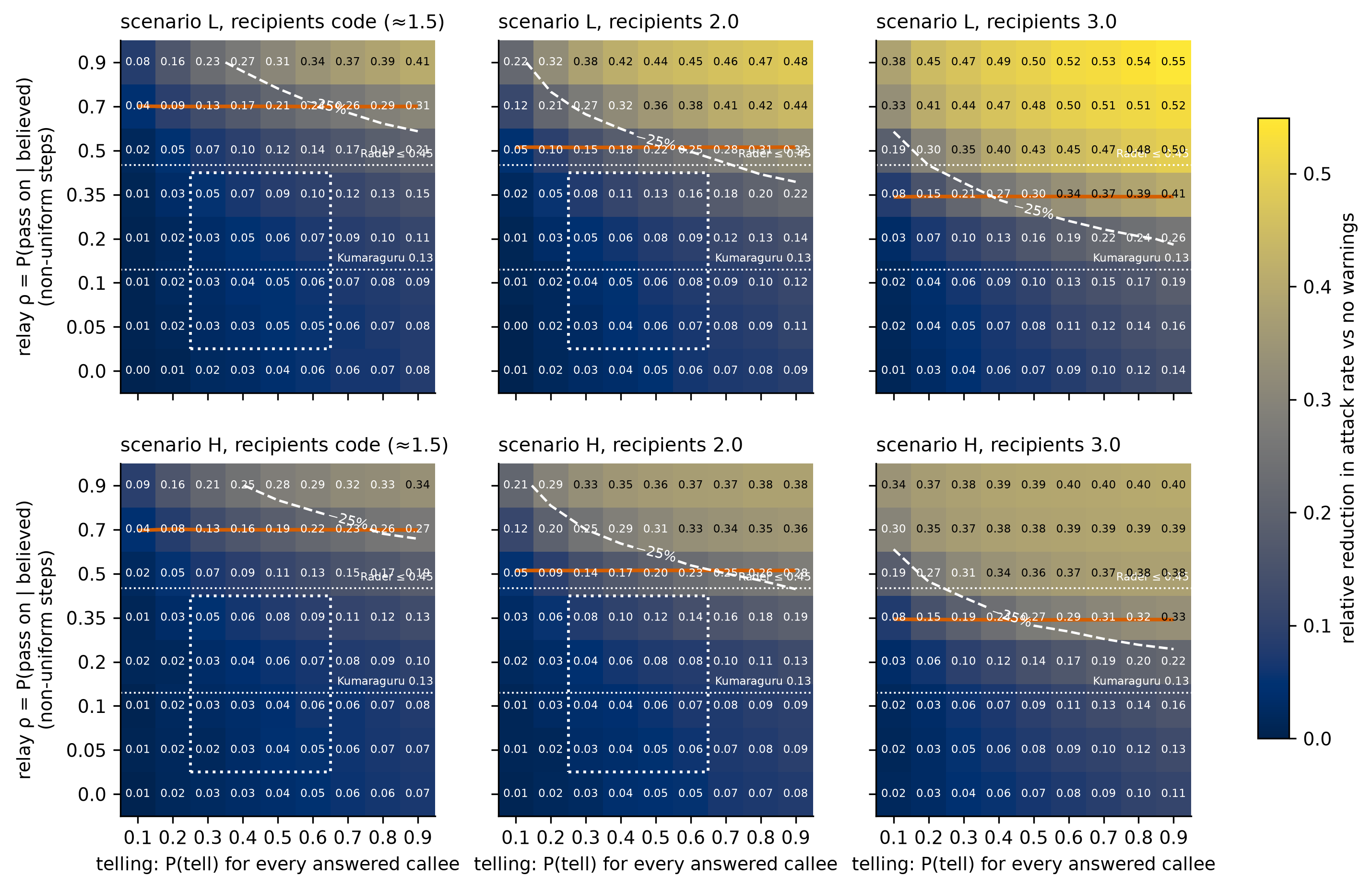}
\caption{\textbf{Full parameter maps with cell values}, relative reduction in the 30-day attack rate, scenarios L (top) and H (bottom). The relay axis has non-uniform steps. Vermilion marks simulated $\Rw = 1$; dashed white the 25\% contour; the dotted box the literature box. 20 graphs $\times$ 25 replicates per cell; Monte Carlo error about $\pm 0.005$.}
\label{fig:phasefull}
\end{figure}

\paragraph{Persistence test.}
\label{app:persist}
Calls stop at day 15 of a literature-centre wave, the generation cap is removed, and warnings are followed to day 75 (Table~\ref{tab:persist}).

Under the main model's message rule, chains persisted exactly when $\Rw \gtrsim 1$, and every cell further above one grew explosively. This persistence is built in: a holder still relays every new delivery, so the process has no depletion. The near-critical message volumes (1.7 and up to 58 believed deliveries per resident per day at days 65--74) are therefore message counts, not people.

Under the new-sender rule, only a delivery that adds a newly held sender is relayed. With engine memory, no chain persisted at $N = 100$, and only rare stragglers persisted near criticality at $N = 1{,}000$. The declared rule anticipated no persistence up to $\Rw \approx 1.5$; with waning, persistence appeared at 1.4--1.5.

\begin{table}[H]
\centering\small
\caption{\textbf{Do warnings persist after calls stop?} Literature centre, scenario L; calls stop at day 15; horizon day 75. Persist means at least one believed delivery in days 65--74, or explosive growth. $N = 100$: 20 live graphs $\times$ 10 replicates; $N = 300$: 5 generated towns $\times$ 4; $N = 1{,}000$: 3 $\times$ 2.}
\label{tab:persist}
\resizebox{\linewidth}{!}{\begin{tabular}{@{}llllll@{}}
\toprule
Told & $\rho$ & $\Rw$ & Message rule, $N = 100$ & New-sender, engine memory, $N$ = 100 / 1,000 & New-sender, 14-d waning, $N$ = 100 / 300 / 1,000 \\
\midrule
2 & 0.2 & 0.40 & 0.00 & 0.00 / 0.00 & 0.00 / 0.00 / 0.00 \\
2 & 0.35 & 0.70 & 0.00 & 0.00 / 0.00 & 0.00 / 0.00 / 0.00 \\
2 & 0.5 & 1.00 & 0.98 & 0.00 / 0.17 & 0.01 / 0.25 / 0.83 \\
2 & 0.7 & 1.40 & 1.00 (explosive) & 0.00 / 0.00 & 0.68 / 1.00 / 1.00 \\
2 & 0.9 & 1.80 & 1.00 (explosive) & 0.00 / 0.00 & 0.88 / 1.00 / 1.00 \\
3 & 0.2 & 0.60 & 0.00 & 0.00 / 0.00 & 0.00 / 0.00 / 0.00 \\
3 & 0.35 & 1.04 & 1.00 & 0.00 / 0.17 & 0.01 / 0.25 / 0.67 \\
3 & 0.5 & 1.50 & 1.00 (explosive) & 0.00 / 0.00 & 0.38 / 0.90 / 1.00 \\
3 & 0.7 & 2.10 & 1.00 (explosive) & 0.00 / 0.00 & 0.86 / 1.00 / 1.00 \\
3 & 0.9 & 2.70 & 1.00 (explosive) & 0.00 / 0.00 & 1.00 / 1.00 / 1.00 \\
\bottomrule
\end{tabular}}
\end{table}

\section{Robustness}
\label{app:robust}
The declared robustness rule called a conclusion robust if its direction held in every arm and its size stayed within a factor of two.

\paragraph{Efficacy curves.} The efficacy curve was the largest source of variation (Table~\ref{tab:curves}). The human-upper curve gave about half the LLM effect at every anchor (ratio 0.48--0.55). The human-lower curve gave 0.40--0.48 of it. Independent doses with the LLM $e_1$ gave 0.068 against 0.085 at the literature centre, so the super-additive step was worth about 20--35\% of the LLM effect. The weaker one's own call protects, the more warnings matter: lowering \eown{} from the in-town value to zero raised the reduction from 0.085 to 0.144.

\paragraph{Broadcast belief.} In the main broadcast results everyone reached believes a trusted source. If only half believe (the institutional level assumed in the live towns, and the primary declared before amendment A12.1), a broadcast reaching 90\% averts 11.8\% alone and 33.8\% with word of mouth (synergy $+7.4$), and a 25\% cut needs 49\% coverage at random and 29\% with residents aged 65+ first; at belief 0.85 the combined effect is 47.1\%. Under 14-day forgetting, the main combined effect falls from 52.1\% to 39.9\%.

\paragraph{Other arms.} Results for the remaining sensitivity arms:
\begin{itemize}[leftmargin=1.2em,itemsep=1pt,topsep=2pt]
\item \emph{Belief.} Source-specific belief levels (0.85, 0.75, 0.60) gave 0.055 (L) and 0.047 (H) at the literature centre, and 0.175 and 0.148 near threshold.
\item \emph{Scam pressure.} Halving or doubling pressure gave 0.055 or 0.132 (L) and 0.035 or 0.131 (H).
\item \emph{Fatigue.} Fatigue was negligible in scenario L and cost 10--17\% in scenario H.
\item \emph{Rewiring.} Degree-preserving rewiring gave 0.082--0.083 (L), and $\Rw$ was 0.38--0.39 on every graph.
\item \emph{Other topologies.} Small-world and scale-free towns gave 0.098 and 0.100 (L), but their mean degree differs.
\end{itemize}

\begin{table}[H]
\centering\small
\caption{\textbf{Efficacy curve $\times$ own-call efficacy.} Relative reduction in the 30-day attack rate at three anchors: engine defaults / literature centre / near threshold (literature with $\rho = 0.35$ and three told). 25 replicates per graph.}
\label{tab:curves}
\resizebox{\linewidth}{!}{\begin{tabular}{@{}lllll@{}}
\toprule
Warning curve & $\eown$ in-town (0.92 / 0.98) & $\eown = 0.5$ & $\eown = 0.15$ & $\eown = 0$ \\
\midrule
LLM panel (isolated-call) & 0.093 / 0.085 / 0.313 & 0.140 / 0.129 / 0.465 & 0.153 / 0.140 / 0.500 & 0.158 / 0.144 / 0.520 \\
LLM in town (first calls) & 0.130 / 0.130 / 0.309 & 0.199 / 0.199 / 0.434 & 0.223 / 0.205 / 0.458 & 0.226 / 0.217 / 0.475 \\
independent doses (LLM $e_1$) & 0.071 / 0.068 / 0.207 & 0.107 / 0.101 / 0.310 & 0.118 / 0.103 / 0.337 & 0.120 / 0.114 / 0.349 \\
human-upper & 0.049 / 0.047 / 0.160 & 0.071 / 0.063 / 0.242 & 0.074 / 0.074 / 0.261 & 0.082 / 0.074 / 0.277 \\
human-lower (independent, human $e_1$) & 0.039 / 0.041 / 0.126 & 0.063 / 0.054 / 0.184 & 0.063 / 0.065 / 0.206 & 0.070 / 0.066 / 0.209 \\
\bottomrule
\end{tabular}}
\end{table}

\begin{figure}[H]
\centering
\includegraphics[width=\linewidth]{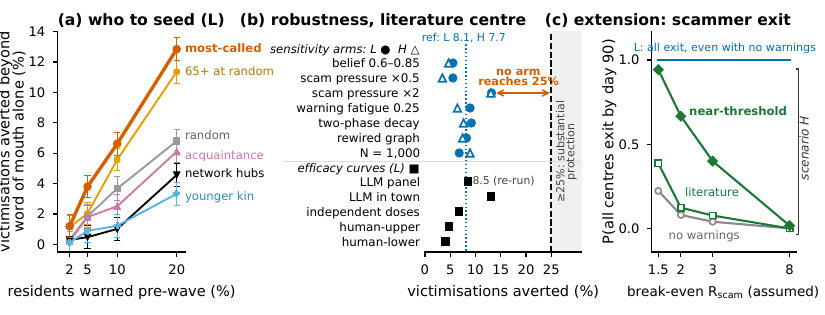}
\caption{\textbf{Education campaigns, robustness and scammer exit} (survey-level settings with 14-day forgetting, 20 consolidation-experiment graphs $\times$ 25 replicates; 95\% Monte Carlo intervals about $\pm$0.7 points). (a) Extra victimisations averted by a pre-wave education campaign giving each seeded resident one tale (scenario L, on top of word of mouth): targeting those scammers call most (12.8\% at 20\% coverage; an oracle, since a real campaign cannot know the lead lists) or residents aged 65+ (11.4\%, its feasible proxy) beats random (6.8\%), hubs (4.6\%) and younger kin (3.3\%). Relay saves about a quarter of the seeds: 10\% most-called seeds with relay match about 13\% without. (b) Victimisations averted in each sensitivity arm (L circles, H triangles) and under five efficacy curves; none reaches the 25\% line (range across all arms 4--22\%). (c) Labelled extension: probability that all four centres exit by day 90 when yield falls below an assumed break-even $R_{\mathrm{scam}}$. In scenario L all exit even without warnings; in H, exit becomes likely only with near-threshold warnings (67--94\% at $R_{\mathrm{scam}} \leq 2$).}
\label{fig:seed}
\end{figure}

\section{Scammer economics (extension)}
\label{app:econ}
Each centre stops dialling for good when its compliance over its last 40 answered calls falls below $\pi^* = \bar c_0/R_{\mathrm{scam}}$. Here $\bar c_0 = 0.244$ is day-0 answered-call compliance and $R_{\mathrm{scam}} \in \{1.5, 2, 3, 8\}$ is assumed. A pattern stays profitable while $R_{\mathrm{scam}}(1 - \rhopop) > 1$, which gives an analogue threshold $\rhopop \geq 1 - 1/R_{\mathrm{scam}}$ \citep{fine2011herd}.

In scenario L, every centre exited within 2--5 weeks with or without warnings, even at $R_{\mathrm{scam}} = 8$; warnings brought exit forward by only 1--10 days. In scenario H, centres rarely exited without warnings (probability 0.00--0.22) or with literature-level warnings (0.00--0.39). Near-threshold warnings, with a 90-day call-weighted protection of about 0.37, made exit likely at $R_{\mathrm{scam}} \leq 2$ (0.67--0.94). Victimisations then fell to 21--28\% of the no-warning, no-exit level, well beyond the 35\% direct reduction. Break-even, the exit rule and the absence of re-entry, fresh leads or new patterns are all assumptions. The extension therefore shows a mechanism by which word of mouth \emph{could} give herd-like protection, through supply rather than transmission \citep{zhu2019investor}, not that it does.

\begin{figure}[H]
\centering
\includegraphics[width=\linewidth]{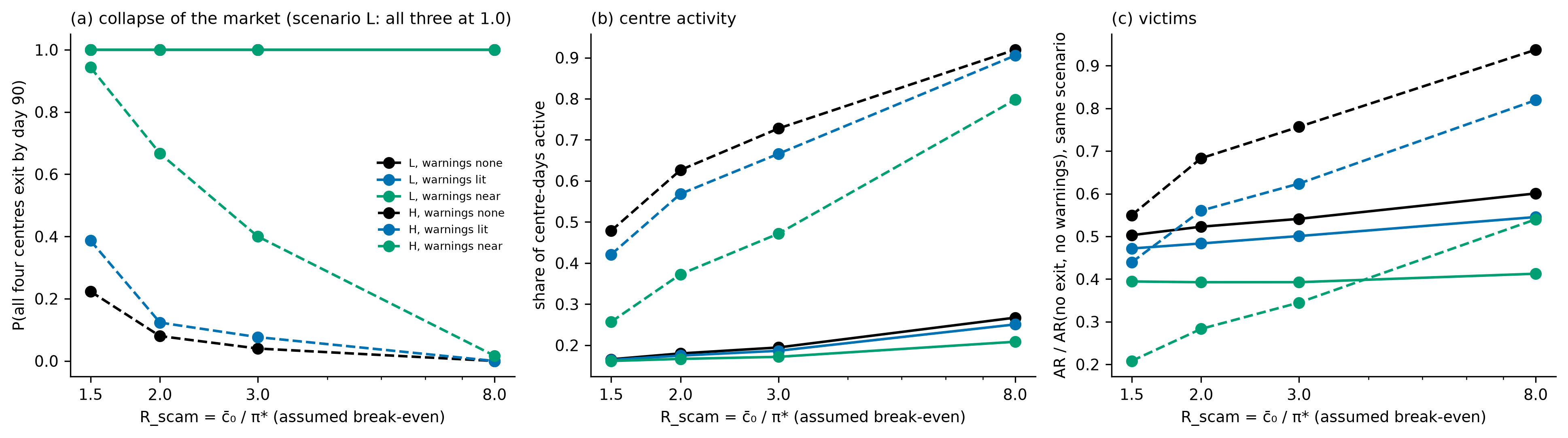}
\caption{\textbf{Scammer economics}, 90 days, 20 graphs $\times$ 15 replicates. (a) Probability that all four centres exit by day 90; (b) share of centre-days active; (c) attack rate relative to no exit and no warnings. Solid lines show scenario L and dashed lines scenario H.}
\label{fig:econ}
\end{figure}

\section{Dose-response panels and control arms}
\label{app:panel}

\paragraph{Panels.} The \emph{isolated-call panel} used 60 residents of one town (30 aged 65+). Each answered the grandparent-scheme call (P02) five times in each of four memory states (1,200 calls):
\begin{itemize}[leftmargin=1.2em,itemsep=1pt,topsep=2pt]
\item naive;
\item one family warning;
\item family plus friend;
\item family, friend and neighbour.
\end{itemize}
Warnings from distinct senders were consolidated into one memory, and the newest was always two days old. Split-half reliability of naive compliance was 0.79.

A \emph{replication panel} (1,800 calls) repeated the naive, one-warning and two-warning states and decomposed the step from one to two warnings ($-0.150$):
\begin{itemize}[leftmargin=1.2em,itemsep=1pt,topsep=2pt]
\item the second warning's text inside the same memory: $-0.100$ (67\%);
\item the ``two people warned you'' count line: $-0.040$ (27\%);
\item vividness wording: $-0.010$ (not resolved).
\end{itemize}

The \emph{control-arm panel} (3,600 calls, \$0.118, none failed) added arms for length, sender, order and wording, with contrasts declared before the run.

\paragraph{Efficacy and super-additivity.} Efficacy is $e_D = 1 - c(D)/c(0)$. Intervals come from a resident-cluster bootstrap (10,000 resamples). Results:
\begin{itemize}[leftmargin=1.2em,itemsep=1pt,topsep=2pt]
\item \emph{Isolated-call panel:} $e_1 = 0.27$ ($-0.08$ to 0.55), $e_2 = 0.88$ (0.73 to 0.98), $S = 0.41$ (0.03 to 1.08), $P(S > 0) = 0.98$.
\item \emph{Replication panel:} naive 0.190, one warning 0.170, two 0.020; $e_1 = 0.105$, $S = 0.70$ (0.22 to 1.69).
\item \emph{Control-arm anchors:} 0.187, 0.123 and 0.037, giving $S = 0.24$ ($-0.09$ to 0.86).
\item \emph{Pooled:} $e_1 = 0.24$ ($-0.08$ to 0.50), $S = 0.44$ (0.09 to 1.05).
\item \emph{Separate memories:} compliance 0.117 against a naive 0.190 ($e_2 = 0.39$). Super-additivity is not resolved against any reference (against the pooled $e_1$: $-0.03$, $-0.40$ to 0.53).
\item \emph{Reversed order (friend, then family):} 0.020 against 0.037, with $S = 0.33$ (0.10 to 0.72).
\item \emph{Neutral wording:} $e_1 = 0.73$ (0.49 to 0.93), and the step from one to two was not resolved ($-0.023$, $-0.063$ to $+0.017$). The declared prediction that neutral wording would weaken both states was wrong.
\item \emph{Same sender:} the same sender twice protected no less than two different senders, so the panel cannot attribute the step to a second person.
\end{itemize}

\paragraph{In-town efficacy.} In the live towns, with town fixed effects in the separate-memory arm, the odds ratios were 0.046 (0.035--0.061) for an earlier own call, 0.28 (0.18--0.43) for one warning and 0.17 (0.07--0.41) for two or more. All panels are exploratory: one pattern, one backbone, residents of one town, and multiple contrasts without adjustment.

\section{Live test: design and full results}
\label{app:live}

\paragraph{Design.} At each of two points, 20 new towns (seeds 700--719) were run twice on the same seed: with disclosure and relay at the point's parameters (ON), and with none (OFF). Both points used the consolidated memory rule, $b = 1$, and 30-day towns of 100 residents. The two co-primary endpoints were:
\begin{itemize}[leftmargin=1.2em,itemsep=1pt,topsep=2pt]
\item ON minus OFF compliance on answered calls, with a paired 95\% $t$-interval (19 degrees of freedom);
\item the relative reduction in attack rate, with a seed-paired percentile bootstrap (10,000 resamples).
\end{itemize}
An effect exists if a Holm-adjusted one-sided test over the four tests rejects ``no reduction''. A forecast is judged:
\begin{itemize}[leftmargin=1.2em,itemsep=1pt,topsep=2pt]
\item \emph{pass} if the interval contains the prediction;
\item \emph{direction only} if the interval excludes zero in the predicted direction but excludes the prediction;
\item \emph{fail} otherwise.
\end{itemize}

The criteria were declared after 51 of the 80 towns had finished and before any result was opened. Three limits were not covered by a declared rule: there was no equivalence bound for P1, no rule for a realism-gate failure, and no measure of memory-slot occupancy.

The forecasts used the in-town table from the consolidation-ON arm. The direction and size class did not depend on the protection form: the OFF-arm table gave $-0.016$ and $-0.033$, and scenario L gave $-0.011$ and $-0.036$. With a paired standard deviation of about 0.019, the P1 compliance endpoint had about 85\% power if the forecast was right. A P1 effect half that size would not have been distinguishable from the OFF$-$OFF gap.

\begin{table}[H]
\centering\scriptsize
\caption{\textbf{Live forecast test: committed predictions, observations and verdicts} (20 seed-paired towns per point, 30 days). $p$ values are one-sided and Holm-adjusted over four tests; attack-rate $p$ values are at the bootstrap resolution floor. Batch-inflated intervals (post hoc, added after the live results were read; A13), which add the batch-to-batch variance estimated from the gap between the two identically configured OFF arms: P1 compliance $-0.0167$ [$-0.0380$, $+0.0046$] and attack-rate reduction 21.0\% [$-3.3$\%, 39.6\%], against P2 $-0.0537$ [$-0.0752$, $-0.0323$] and 60.6\% [48.7\%, 69.7\%]; the batch variance rests on a single OFF--OFF replicate.}
\label{tab:ee}
\setlength{\tabcolsep}{3pt}\begin{tabularx}{\linewidth}{@{}L>{\raggedright\arraybackslash}p{1.2cm}>{\raggedright\arraybackslash}p{2.3cm}>{\raggedright\arraybackslash}p{1.0cm}>{\raggedright\arraybackslash}p{1.2cm}>{\raggedright\arraybackslash}p{2.3cm}>{\raggedright\arraybackslash}p{1.3cm}@{}}
\toprule
Quantity & P1 predicted & P1 observed & P1 verdict & P2 predicted & P2 observed & P2 verdict \\
\midrule
\textbf{ON $-$ OFF compliance} & $\mathbf{-0.014}$ & $\mathbf{-0.0167}$ [$-0.0254$, $-0.0080$], $p = 0.0004$ & \textbf{pass} & $\mathbf{-0.033}$ & $\mathbf{-0.0537}$ [$-0.0629$, $-0.0446$], $p < 10^{-9}$ & \textbf{direction only, larger} \\
\textbf{Relative AR reduction} & \textbf{19\%} & \textbf{21.0\%} [11.4\%, 29.4\%], $p \leq 0.0003$ & \textbf{pass} & \textbf{47\%} & \textbf{60.6\%} [56.4\%, 64.4\%], $p \leq 0.0003$ & \textbf{direction only, larger} \\
Compliance, OFF arm & 0.071 & 0.077 & -- & 0.071 & 0.087 & -- \\
Compliance, ON arm & 0.057 & 0.060 & -- & 0.038 & 0.033 & -- \\
AR, OFF $\rightarrow$ ON & 0.235 $\rightarrow$ 0.190 & 0.250 $\rightarrow$ 0.198 & -- & 0.235 $\rightarrow$ 0.125 & 0.283 $\rightarrow$ 0.112 & -- \\
Ever-victim share, OFF $\rightarrow$ ON & 0.211 $\rightarrow$ 0.175 & 0.214 $\rightarrow$ 0.169 & -- & 0.211 $\rightarrow$ 0.120 & 0.235 $\rightarrow$ 0.102 & -- \\
\phitwo{} (dialled) & 0.17 & 0.175 & -- & 0.70 & 0.708 & -- \\
Depth 0 / 1 / 2 / 3+ (ON) & 0.58 / 0.25 / 0.10 / 0.07 & 0.58 / 0.24 / 0.10 / 0.08 & -- & 0.21 / 0.09 / 0.07 / 0.63 & 0.21 / 0.08 / 0.06 / 0.64 & -- \\
WBS, all / 65+ & 0.29 / 0.26 & 0.274 / 0.252 & -- & 0.63 / 0.62 & 0.625 / 0.605 & -- \\
$\Rw$ (generation ratio) & 0.38 & 0.38 & -- & 1.17 & 1.17 (6-gen.\ cap) & -- \\
Believed deliveries per town & $\approx$550 & 543 & -- & $\approx$7,300 & 7,450 & -- \\
Realism gates G1--G4 (ON / OFF) & -- & pass / pass & -- & -- & pass / pass & -- \\
\bottomrule
\end{tabularx}
\end{table}

\paragraph{Secondary results.}
\begin{itemize}[leftmargin=1.2em,itemsep=1pt,topsep=2pt]
\item \emph{Absolute effect.} Warnings removed 0.053 victimisations per resident over 30 days at P1 (0.024 to 0.081) and 0.172 at P2 (0.143 to 0.200).
\item \emph{By age.} At P2, compliance among residents aged 65+ fell from 0.117 to 0.047.
\item \emph{Relay.} Believed deliveries by generation, summed over towns, rose at P2 from 16,374 first-hand to 35,558 in the sixth generation, so relay there was supercritical and held finite only by the cap.
\item \emph{Deep warnings.} Residents aged 65+ holding four or more warnings complied on none of 2,298 answered calls. This figure is descriptive and confounded: deep-warned calls come late in the wave, after own-call immunity has set in.
\end{itemize}

\paragraph{The P2 miss.} The forecast ON$-$OFF compliance was $0.038 - 0.071 = -0.033$, and the observed was $0.033 - 0.087 = -0.054$ (means of per-town rates). Of the $-0.021$ gap, about $-0.016$ is the OFF arm above its forecast, which the depth cap cannot explain. About $-0.005$ is the ON arm below its forecast.

\paragraph{Noise check and operations.} The two OFF arms share one configuration. A mock run of the two gave 3,168 byte-identical requests, and every live request was served by the pinned provider. Yet live compliance in P2 OFF exceeded P1 OFF by $+0.010$ ($+0.004$ to $+0.016$; paired $p = 0.003$; 16 of 20 pairs positive), and the attack rate by $+0.033$.

An undeclared sensitivity analysis compared each ON arm with the per-seed mean of both OFF arms:
\begin{itemize}[leftmargin=1.2em,itemsep=1pt,topsep=2pt]
\item P1: $-0.0217$ ($-0.0297$ to $-0.0138$) and 25.9\% (18.0--32.9\%), both pass;
\item P2: $-0.0487$ and 58.2\% (53.3--62.5\%), both direction only.
\end{itemize}

Eight of the 80 towns stopped mid-run on a truncated provider response that the client did not retry. They were resumed from per-tick checkpoints with the same seeds and configurations, and the engine restored the full simulation state, including random-number streams. The resume ran from 19:53 to 20:00 and the results commit is timed 20:04, so the decision preceded any reading of results. No seed was replaced and no pair was dropped. The client now retries such errors.

\section{Friction and the consolidation test}
\label{app:friction}
Friction, the refusal of legitimate calls, did not rise measurably with warnings in the live towns. In the separate-memory arm, residents refused 0.072 (0.061--0.085) of legitimate calls with no warnings, 0.063 with one warning and 0.057 with two or more. With consolidated memories, refusal rose slightly to 0.085 at two or more warnings.

Discrimination rose with dose, from $d' = 2.0$ to 4.2, almost entirely through scam resistance. The legitimate calls are easy to tell from scams by design, so this is not evidence that friction is absent in people, for whom generic tips raise false refusals \citep{kubilay2023can}.

The declared consolidation test, in 20 paired 30-day towns, found an effect of $-0.0057$ ($-0.0145$ to $+0.0031$; $p = 0.19$). This is in the predicted direction but not supported. The ratio of observed to the panel-based forecast was 0.52 ($-0.28$ to 1.33), compatible both with the forecast and with no effect. Its precondition was forecast accurately: 10.3\% of received calls predicted to reach a twice-warned resident, against 10.4\% observed.

\section{Declaration and amendment trail}
\label{app:trail}
Most stages were declared before their results were read, with their design, primary endpoint and decision rule, in the project's version-controlled results log. The exceptions, each noted where it arises, are: the live-test pass/fail criteria, committed after 51 of 80 towns had finished (the forecasts themselves were committed before any live result; Appendix~\ref{app:live}); the broadcast-belief amendment A12.1; the post-hoc batch inflation A13; and an undeclared comparison against the mean of both OFF arms. After the first broadcast run (A12, primary belief 0.50) had been read, amendment A12.1 made belief 1.0 the main case by a PI decision (a trusted source is believed, as personal ties are), before any belief-1.0 computation; belief 0.50 and 0.85 remain sensitivities. These are internal declarations in a git repository, not public registry entries. The live-test declaration, predictions, criteria and results will be deposited publicly with their commit hashes.

Stages A1--A4 and the first transmission test ran on MiniMax-M3; everything from A5 on ran on Qwen3.7 Flash, and results from the two periods are never pooled. Some declarations were not blind, and the log says so:
\begin{itemize}[leftmargin=1.2em,itemsep=1pt,topsep=2pt]
\item the A4 backbone rule reached the analyst after the probes had been read;
\item the A11 simulator was designed after a descriptive look at the consolidation logs;
\item the round-1 super-additivity interval re-analyses data that had already been read.
\end{itemize}

Three analyses are confirmatory, with declared decision rules: the consolidation experiment and the two points of the live test. Everything else is exploratory, or a model output with declared predictions.

\begin{table}[H]
\centering\scriptsize
\caption{\textbf{Declaration and amendment trail}, compact (git commit dates, 2026; full log in the repository).}
\label{tab:trail}
\begin{tabularx}{\linewidth}{@{}p{1.5cm}p{1.1cm}LL@{}}
\toprule
Stage & Date & Declared & Outcome \\
\midrule
Stage 1--A3 & 10-10/11 & first transmission test (MiniMax-M3); calibration to real-world rates & pipeline check; calibration failed, real-rate matching dropped \\
A4 & 10-11 & realistic call scripts; realism gates; backbone screen & gates pass; Qwen3.7 Flash selected (rule not blind) \\
A5--A7.1 & 10-11/12 & belief-probe transmission tests; LLM-decided vs modelled disclosure and uptake & not supported; LLM social behaviour unrealistic, so modelled \\
A8 & 10-12 & personal ties always believed & design change \\
A9, A9.5 & 10-12 & isolated-call dose-response panel and replication & 0.197 / 0.143 / 0.023 / 0.007 (exploratory) \\
A10, A10.1 & 10-12 & live consolidation test (confirmatory, 20 pairs) & $-0.0057$ [$-0.0145$, $+0.0031$], not supported; reach forecast met \\
A11 & 10-12 & fast simulator, fidelity gate, maps, seeding, robustness, economics & gate passes; 8.6\% (L), 7.4\% (H) \\
Live test & 10-12 & ON vs OFF at P1 and P2, 20 pairs each; forecasts committed before any live result, pass/fail criteria after 51 of 80 towns (a27c827, 15eb814, ff17b81) & P1 pass; P2 direction only; not forecast-validated (89f1ff5) \\
A11.R1 & 10-12 & control-arm panel; persistence test; reconciliation runs & exploratory (1ebca6a, f0cd36e) \\
A12 & 10-12 & broadcast: coverage, word of mouth, two doses; primary belief 0.50 & 90\%: 11.8\% alone, 33.8\% with word of mouth (L) \\
A12.1 & 10-13 & amendment after A12 was read: belief 1.0 becomes the main case (PI decision) & 90\%: 23.4\% / 52.1\%; synergy $+14.2$ (L), $-5.5$ (in-town) \\
A13 & 10-13 & batch-inflated live intervals; $R_{\mathrm{ngm}}$ check (post hoc, after the live results were read; 394faa2) & P1 intervals include zero, P2's do not; $R_{\mathrm{ngm}} \approx R_{\mathrm{branch}}$ (4dafffd) \\
\bottomrule
\end{tabularx}
\end{table}

\section{What we tried, and LLM methods lessons}
\label{app:lessons}
The design in the main text is the product of declared, reported failures.
\begin{itemize}[leftmargin=1.2em,itemsep=1pt,topsep=2pt]
\item \emph{Stimulus realism over calibration.} Early scripts let legitimate calls carry scam tactics, and warned residents refused 59 of 60 legitimate calls. Two declared recalibrations to real-world rates failed held-out validation. Realistic scripts from 48 sources raised warned legitimate acceptance to 0.50 with warned scam compliance at zero, and realism gates replaced rate targets.
\item \emph{Backbones differ.} Of five backbone and endpoint combinations screened, two passed the gates; DeepSeek V4 Flash refused 75\% of legitimate calls after a warning. All results use Qwen3.7 Flash.
\item \emph{Stated beliefs overstate behaviour.} A warning moved stated compliance from 0.444 to 0.215 while actual compliance moved from 6 to 5 of 40; probe-based protection was about six times the behavioural estimate.
\item \emph{LLM uptake rubber-stamps warnings.} Qwen believed 99.8\% of 8,420 warnings and would relay 99.7\%, so belief and relay are modelled.
\item \emph{LLM victims are not silent.} 82--93\% of LLM victims said they would tell someone, against about 0.15 in the evidence-based model, so disclosure is modelled.
\end{itemize}

\section{LLM requests and prompts}
\label{app:prompts}
The backbone was \texttt{qwen/qwen3.7-flash} via OpenRouter, with the provider pinned to Alibaba and no fallbacks. Reasoning was disabled, the temperature was not sent (so the provider default applied, which the provider does not report), and output was JSON.

\paragraph{System frame.} ``This is a research simulation of a fictional town, used to study how people protect each other from phone fraud. You play one fictional resident. Answer as that person realistically would, with their habits, worries and blind spots, not as an AI assistant and not as a fraud expert. Real people are often trusting, distracted, polite and inconsistent. Return only the requested JSON.''

\paragraph{Persona.} ``You are \textit{name}, aged \textit{age}. You live alone. Your comfort with technology is low. When you are fooled you are quite embarrassed, and you confide sometimes, mostly to your child. Your trust in banks and agencies is moderate, and you are sometimes lonely.''

\paragraph{Memories.} Up to 12, one per line. Examples:
\begin{itemize}[leftmargin=1.2em,itemsep=1pt,topsep=2pt]
\item single warning: ``[2 days ago, from your family contact, clear, you fully believe it] Warning: \textit{text}'';
\item consolidated warning: ``[2 people you know warned you about this, most recently 2 days ago, vivid, you fully believe it] Warning: \textit{newest text} An earlier warning: \textit{first text}'';
\item neutral wording: ``[2 days ago, from your family contact] Message: \textit{text}''.
\end{itemize}

\paragraph{\texttt{call\_response}.} ``\textit{memories}. Your phone rings. Caller ID shows: \textit{caller id}. If you answer: \textit{script}. What do you do? action: not\_answer, comply, resist, hang\_up, or verify (end the call and check it yourself through a number or channel you already know). suspicion: 0 to 1. utterance: one short sentence you would say.''

\paragraph{\texttt{after\_call}.} ``Earlier today you got a phone call: \textit{script}. What happened: \textit{outcome}. takeaway: what, if anything, you yourself take away from this \ldots{} message: if you did tell someone about it, the short message you would actually say.'' Whether the message is sent is decided by the disclosure model.

\section{Compute and cost}
\label{app:cost}
All closed-loop analyses ran on CPU at no LLM cost. The fidelity gate used 8,000 replicates of about 0.015 s each on 12 workers, and the parameter maps simulated 249,000 town-months in 34 minutes on 12 workers.

Recorded LLM spend over the whole project was about \$24.19 on MiniMax-M3, mostly on the failed calibrations, and about \$24.94 on OpenRouter through the consolidation experiment. Of the OpenRouter spend, the measurements used here cost \$0.038 (isolated-call panel), \$0.059 (replication panel) and \$6.40 (40 live consolidation towns). The control-arm panel added \$0.118. The live forecast test cost \$13.83 by the key's usage (budget \$15): 195,541 requests, all served by the pinned provider, over 2 h 48 min with 48 towns in parallel, plus a \$0.31 resume. Total recorded LLM spend for the paper is about \$63.

\end{document}